\RequirePackage{fix-cm}
\documentclass[twocolumn]{svjour3}          
\smartqed  
\usepackage[numbers,sort&compress]{natbib}
\usepackage{amssymb, amsfonts, amsmath, graphicx, subfigure, bbding, lipsum, multirow, makecell, picinpar, multirow, enumerate}
\usepackage[linesnumbered,  ruled]{algorithm2e}
\begin{document}

\title{Connectivity-preserving Geometry Images}

\author{Shaofan Wang \and  Dehui Kong \and Juan Xue \and Weijia Zhu \and Min Xu \and Baocai Yin \and Hubert Roth
}
\institute{S. Wang, D. Kong, J. Xue, W. Zhu, B. Yin  \at
Beijing Key Laboratory of Multimedia Intelligent Software Technology, College of Metropolitan Transportation,
Beijing University of Technology, Beijing 100124, China \\
              \email{wangshaofan@bjut.edu.cn}           
           \and
           M. Xu  \at
           School of Mathematical Sciences, Dalian University of Technology, Dalian 116085, China
           \and
           H. Roth \at Institute of Automatic Control Engineering, University of Siegen, Siegen 57068, Germany
}

\date{Received: date / Accepted: date}

\maketitle

\begin{abstract}

We propose \emph{connectivity-preserving geometry images} (CGIMs), which map a triangular mesh onto a rectangular regular array of an image, such that the reconstructed mesh produces no sampling errors, but merely round-off errors over the coordinates of vertices. By using permutation techniques on vertices,
CGIMs first obtain a \emph{V-matrix} whose elements are vertices of the original mesh, which intrinsically preserves the vertex-set and connectivity of the original mesh, and then generate a CGIM array by transforming the Cartesian coordinates of corresponding vertices of the V-matrix into RGB values.
Compared with traditional geometry images (GIMs), CGIMs achieve the minimum reconstruction error with a parametrization-free algorithm. We apply CGIMs to lossy compression of meshes. Experimental results show that while CGIMs produce a lower efficiency in both encoding and decoding time and larger resolutions than traditional GIMs,
CGIMs perform better Peak Signal-to-Noise Ratios and preserve details better than GIMs especially with the multi-stage base color and index map scheme, because CGIMs treat details and non-details of meshes evenly as all elements of the V-matrix.
\end{abstract}

\section{Introduction}

Geometry images (GIMs) are a completely regular remeshing method, which represents a three-dimensional mesh using an image-like structure, with the connectivity information encoded in the image space \cite{gim-02}. The main procedure of GIMs includes three steps: mesh parametrization, sampling and scaling. A three-dimensional mesh is first mapped onto a square or spherical parametric domain, and a three-dimensional matrix is then obtained by sampling coordinates of points over the parametrized domain, and the matrix is finally transformed into an image array with a proper scaling. To rebuild the mesh, the vertices are obtained by transforming the RGB values of the image into Cartesian coordinates, and the edges are obtained by connecting all the pairwise vertices which are adjacent in the array.

The reconstruction error arising from GIMs includes two parts: the \emph{sampling error} produced in the sampling step, depending on the parametrization methods, the sampling methods, GIM resolutions and interpolating functions, and the \emph{round-off error} produced trivially in the scaling step, depending on the scaling between the coordinate values of vertices and RGB values of GIMs.

To effectively decrease reconstruction errors, a solution is to adopt a mesh parametrization minimizing the geometric stretch in GIMs, where the stretch metric is derived from a Taylor expansion of geometric errors. Although such a parametrization is directly derived from the reconstruction error, it is difficult to achieve the global minimizer of the metric because of the multivariate nonconvex optimization problem. An alternative is to first map the whole mesh piecewise onto several charts, each of which has small distortion and hence is effectively sampled, and then pack them together into a GIM. Such methods, referred to as multi-chart GIMs
\cite{mulcht-03,rectangular-06,adagim-08,featurepre-2010}, not only solve the large distortion problem but also handle genus-nonzero meshes. However, multi-chart GIMs have low efficiency in packing irregular charts, as the graph cut has to be performed to fit both decreasing geometric stretch over a mesh and fewer cutting nodes and branches.

This paper proposes \emph{connectivity-preserving geometry images} (CGIMs), which map a three-dimensional triangular mesh onto an image array which intrinsically preserves the vertex-set and the connectivity of the original mesh. The motivation comes from the observation that the connectivity of the mesh be preserved by allowing each vertex to appear repeatedly in an image array (the second column of Fig.~\ref{fig-igim1}). Such a structure of CGIMs treats every vertex evenly which preserves details of meshes well. Compared with traditional GIMs, CGIMs produce larger resolutions for representing meshes and spend more time for encoding CGIM arrays and reconstruction from compressed CGIM arrays. However,  experimental results of lossy compression show that by using the  multi-stage base color and index map (MBCIM) scheme as image codec, CGIMs give promising results in both detail preservations and the Peak Signal-to-Noise Ratio (PSNR) curves.

The paper is organized as follows. We introduce related work regarding GIMs in Section~\ref{sec-2}. We give an overview of CGIMs in Section~\ref{sec-3}. We show the main results of this paper in Section~\ref{sec-4}, where we give the main idea of all the phases of CGIMs, a CGIM property and a complete CGIM algorithm. We give other CGIM algorithms in Section~\ref{sec-5}, including CGIMs with smaller resolutions and mesh reconstruction from compressed CGIMs. We apply CGIMs to lossy compression of meshes in Section~\ref{sec-6} by using JPEG2000 and the MBCIM scheme. We conclude this paper in Section~\ref{sec-7} with limitations of CGIMs and future work.

\section{Related Work} \label{sec-2}

\emph{Single-chart GIMs}\quad
The pioneering work of GIMs  performs a graph-cut over closed meshes, maps them onto a square domain
by using a minimizing-geometric-stretch parametrization,
and imposes a regular sampling for surface geometry \cite{gim-02}.
Zhou et al. present GIMs using adaptive sampling, and employs the JPEG2000 codec
for mesh compression \cite{zhou-04}. Praun and Hoppe  use spherical parametrization to map a genus-zero surface
onto a spherical domain \cite{sph-03}. Gauthier and Poulin  propose another spherical GIM
to treat arbitrary genus surfaces, with the holes explicitly represented using genus reduction \cite{arbigenus-09}.
Meng et al.  integrate differential coordinates into traditional GIMs for geometric morphing \cite{diff-10}.

\emph{Multi-chart GIMs}\quad
As mapping an entire surface to a single chart may produce large distortion, multi-chart GIMs are developed.
Sander et al. map the mesh piecewise onto several charts and pack them together into a GIM \cite{mulcht-03}.
Carr et al.  partition a mesh into quasi-rectangular clusters which map to rectangular charts in parameter space \cite{rectangular-06}.
Yao and Lee  decompose a mesh into square GIM charts with different resolutions, each of which is adaptively determined by
a local reconstruction error \cite{adagim-08}. Feng et al. generate triangular patches
for input meshes using a curvilinear feature that preserves salient features and supports GPU-based LOD representation of meshes \cite{featurepre-2010}.

\emph{Applications of GIMs}\quad
Research work on applications of GIMs includes mesh compression \cite{shape-05}, smooth surface representation \cite{smoo-03},
face recognition \cite{intraclass-07}, texture synthesis \cite{textsyn-05}, and facial expression modeling \cite{modelface-12}.
We do not list detailed work in order to focus in priority on our work.

\section{An Overview of CGIMs} \label{sec-3}

The main idea of CGIMs is to arrange all the vertices of a mesh into a \emph{V-matrix} (i.e. a matrix whose elements uniquely correspond to vertices of the mesh) by inserting them  repeatedly, such that the vertex-set and the edge-set generated by the V-matrix are equal to the vertex-set and the connectivity of the original mesh respectively.
Then a CGIM array is obtained by transforming the Cartesian coordinates of all elements of the V-matrix to pixel values.
To reconstruct a mesh from the CGIM array, we obtain the vertex-set by collecting elements with different encoding coordinates,
and obtain the edge-set by collecting pairwise elements of the array with different encoding connectivity.

Accordingly, a CGIM algorithm contains three phases: \textsf{isomatrix}, \textsf{isogim} and \textsf{reconstruct},
where the first phase gives a V-matrix, the second one transforms the V-matrix into a CGIM array,
and the last one reconstructs a mesh from the CGIM array. Among the three phases, \textsf{isomatrix} is key for CGIMs, whose algorithm is accordingly tricky and complicated.
Before we detail  its algorithm in the next section, we shall give a simple understanding of the main idea of \textsf{isomatrix}. The \textsf{isomatrix} phase includes the following three sub-phases. The first phase partitions the vertex-set of the mesh into several ordered subsets called \emph{levels}, the second phase adds repeated elements within those levels so that each pair of neighboring levels preserves an associated edge-set,
and the last phase adds repeated elements within the levels again to maintain them in an image-like structure. The rule for adding repeated elements is based on the edge-set induced by each pair of neighboring levels.

\begin{figure}[t] \centering
\includegraphics[width=0.49\textwidth]{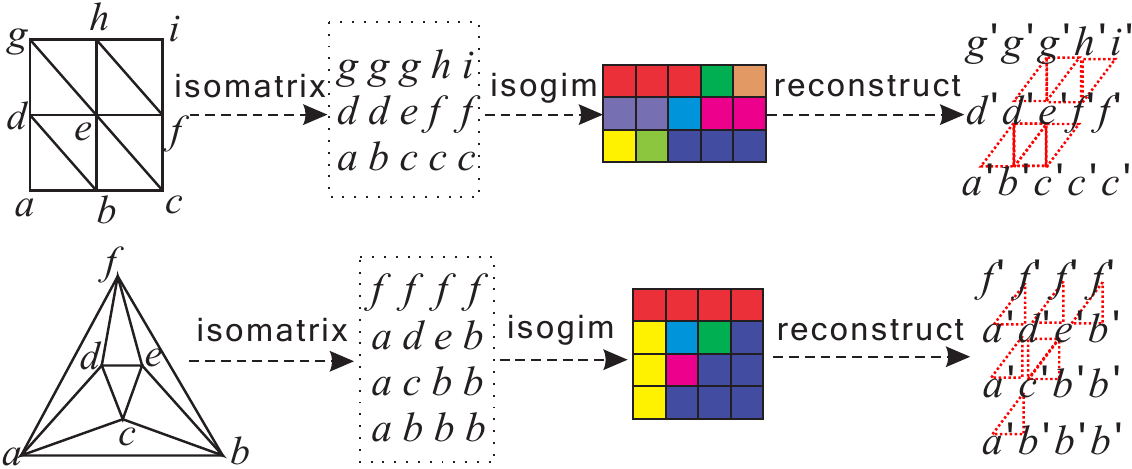}
\caption{CGIM examples. Columns 1-4: original meshes, V-matrices, CGIM arrays, reconstructed meshes, respectively.}\label{fig-igim1}
\end{figure}
In the following introduction of three sub-phases, we define a \emph{level} of length $n$ to be an $n$-tuple $[v_1,\ldots,v_n]$ whose elements are vertices of the mesh, and call each  $v_i$ to be an \emph{element} of the level.

\textbf{\textsf{Stratify}} \quad The \textsf{Stratify} phase gives a sequence of levels of the mesh (each level contains no repeated elements but different levels may share common elements and may have different lengths) such that the union of all elements of all the levels is equal to the vertex-set of the mesh.
Obtaining such a level sequence is a fundamental work for the \textsf{isomatrix} phase, and has to satisfy several properties described in Lemma~1, Section~A of the supplementary material. Basically such properties guarantee that each edge of the mesh be characterized either by pairwise neighboring elements of a level (such an edge is referred to as an \emph{intra-level edge}), or by pairwise elements from two neighboring levels (such an edge is referred to as an \emph{inter-level edge})
\footnote{
 The \textsf{Stratify} phase essentially divides all faces of the input mesh into several triangle strips \cite{tristrip-07}, i.e., series of connected triangles sharing vertices. To see this, we observe that the first example of Fig.~\ref{fig-igim1} can be described by using two strips: $dgehfi, adbecf$, with $dgehfi$ corresponding to the face set generated by the first and second output levels, and $adbecf$ corresponding to the face set generated by the second and third output levels (the red triangles of the last column).
 Similarly, the second example of Fig.~\ref{fig-igim1} can be described by using three strips: $afdefb, adceb, acb$,  with $afdefb$ corresponding to the face set generated by the first and second output levels, and $adceb$ corresponding to the face set generated by the second and third output levels, and $acb$ corresponding to the face set generated by the third and fourth output levels. While triangle strips propose an efficient storage of triangular meshes, such a method cannot generate an image-like array of vertices and hence no image codec can be imposed.}.

\textbf{\textsf{Align2Levels}} \quad  The \textsf{Align2Levels} phase transforms each level obtained in previous phase to two levels, one of which coincides with the former level and the other of which coincides with the latter level, by inserting repeated elements within each level. This work guarantees that the lengths of two neighboring levels be the same and the end-vertices of each inter-level edge appear in correct locations.

\textbf{\textsf{AlignAllLevels}} \quad  Although the inter-level edges are preserved in pairwise levels and the intra-level edges are preserved in neighboring elements of each level, we cannot obtain a V-matrix preserving the vertex-set and the edge-set of the mesh unless all the levels obtained from previous phase have the same length. The \textsf{AlignAllLevels} phase accomplishes this work by adding repeated elements within each level such that the pairwise levels obtained from previous phase are exactly the same.

\section{Details of CGIMs} \label{sec-4}

This section gives details of all phases of CGIMs. We first introduce some definitions and notations.

Let $\mathcal{M}$ be a triangular mesh in $\mathbb{R}^3$.
We denote $v\sim w$ if the vertices $v$, $w$ are adjacent in $\mathcal{M}$, otherwise we denote $v\not\sim w$;
we denote $v= w$ if the vertices $v$, $w$ are the same vertex in $\mathcal{M}$, otherwise we denote $v\neq w$.
We denote $\boldsymbol{V}$, $\boldsymbol{E}$, $\boldsymbol{V_B}$ to be the vertex-set, the edge-set,
the collection of all boundary vertices of $\mathcal{M}$ respectively,
and denote $\boldsymbol{N}(v)$ to be the collection of all the vertices adjacent to $v$ in $\mathcal{M}$.

A \emph{V-matrix} with respect to $\mathcal{M}$ is defined to be a two-dimensional array whose elements
are vertices of $\mathcal{M}$;  a \emph{level} of $\mathcal{M}$ is defined to be an $n$-tuple $[v_1,\ldots,v_n]$ whose elements are vertices of $\mathcal{M}$,
where $n$ is called the length of the level. We denote  $\mathcal{L}(j)$ to be the $j$-th element of a level $\mathcal{L}$,
and we call $j$  to be the \emph{index} of $v$ in $\mathcal{L}$ if $v=\mathcal{L}(j)$, $1\leq j\leq n$.
We denote $\mathcal{L}({\rm end})$ to be the last element of $\mathcal{L}$,
and denote $\mathcal{L}(j_1:j_2):=[\mathcal{L}(j_1), \mathcal{L}(j_1+1),\ldots,\mathcal{L}(j_2)]$ if $j_1\leq j_2$;
otherwise, we denote $\mathcal{L}(j_1:j_2):=\varnothing$. We denote $|A|$ to be the number of all elements of a set $A$
or the length of a level $A$, and denote $|\mathcal{L}|_{v}$ to be the number of all $v$'s of a level $\mathcal{L}$.
We adopt the convention that when we perform set operations (e.g. $\in,\cap,\cup,\setminus$) on a level $\mathcal{L}$,
then $\mathcal{L}$ represents the collection of all different elements of the level $\mathcal{L}$. 
Let $v,v'$ be vertices and let $\mathcal{L},\mathcal{L}'$ be levels. Then $v$ is called an adjacent element of $v'$ in $\mathcal{L}$ if $v'\sim v\in\mathcal{L}$; moreover, $v$ is called an adjacent element of $\mathcal{L}'$ in $\mathcal{L}$ if  there exists $v'\in\mathcal{L}'$ such that $v'\sim v\in\mathcal{L}$.

\begin{window}[3, r, {\mbox{
\includegraphics[width=0.25\textwidth]{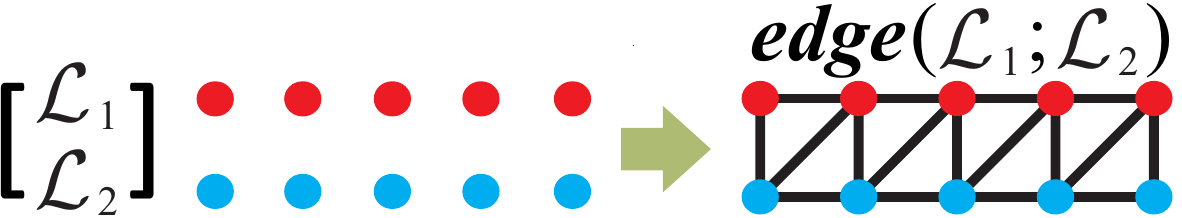}}},{}]
Let $\mathcal{L}_1, \mathcal{L}_2$ be two levels of $\mathcal{M}$ with the same length.
The \emph{edge-set induced by} $\mathcal{L}_1, \mathcal{L}_2$, denoted by $\boldsymbol{edge}(\mathcal{L}_1; \mathcal{L}_2)$, is defined to be the collection of pairwise vertices obtained by connecting each pair of distinct neighboring elements (the horizontal, vertical and slash directions) of $\mathcal{L}_1, \mathcal{L}_2$, i.e.
\end{window}
\vspace{-5mm}
\begin{align*}
 \boldsymbol{edge}(\mathcal{L}_1; \mathcal{L}_2) := \cup_{j} &\{\mathcal{L}_1(j)\mathcal{L}_1(j\!+\!1),~\mathcal{L}_2(j)\mathcal{L}_2(j\!+\!1), \\
&    ~~\mathcal{L}_1(j)\mathcal{L}_2(j), ~\mathcal{L}_1(j\!+\!1)\mathcal{L}_2(j)\},
\end{align*}
each element of which has distinct end-vertices, with the index $j$  taking all available values.

Let $\mathcal{L}_j, j=1,\ldots,n$ be a level sequence of the mesh, $n\geq3$. The \textbf{concatenate} operator $\langle\cdot\rangle$ which maps the level sequence to a novel level is deductively defined by
\begin{align*}  &  \langle \mathcal{L}_1,\mathcal{L}_2\rangle:= [\mathcal{L}_1(1),\ldots,\mathcal{L}_1({\rm end}),~ \mathcal{L}_2(1),\ldots,\mathcal{L}_2({\rm end})],  \\
& \langle \mathcal{L}_1,\mathcal{L}_2,\ldots,\mathcal{L}_n \rangle := \langle ~ \langle \mathcal{L}_1,\mathcal{L}_2,\ldots,\mathcal{L}_{n-1}\rangle, ~ \mathcal{L}_n \rangle. \end{align*}

The following definitions are given for partitioning the vertex-set into several levels of the mesh; such partition is non-trivial as we can see the difference from two examples of Fig.~\ref{fig-igim1} (in the first example the vertices are ``well organized" so that the partitioned levels share no common vertices; however in the second example, the partitioned levels share some vertices in order to preserve the connectivity of the mesh). The \textbf{irregular pair} in the following definition is an important tool to check whether the vertices are ``well organized".

\begin{definition}\label{def-component}
Let $Q:=[q_1,q_2,\ldots,q_n]$ be a level of a mesh.  A \textbf{sublevel} $[q_{i_1},q_{i_2},\ldots,q_{i_k}]$, $1\leq k\leq n$ of $Q$ is defined to be a level consisting of elements of $Q$ obeying the same relationship of indices of $Q$, i.e. $1\leq i_1<i_2<\cdots<i_k\leq n$.
A \textbf{component} $[q_{i_1},q_{i_1+1},\ldots,q_{i_2-1},q_{i_2}]$ of $Q$ is defined to be a sublevel of $Q$ such that
\[ q_{i_1-1}\not\sim q_{i_1}\sim q_{i_1+1} \sim \cdots \sim q_{i_2-1} \sim q_{i_2}\not\sim q_{i_2+1} \]
with $1\leq i_1\leq i_2\leq n$,  where the first (the last, respectively) relationship is valid when the subscript of $q$ arrives to 0 ($n+1$, respectively). A set of single vertex $\{q_{i}\}$ is defined to be a component of $Q$ if $q_{i}\not\sim q_{i-1}$ and $q_{i}\not\sim q_{i+1}$.
\end{definition}

\begin{definition}\label{def-irre}
Let $Q^{\emph{c}}:=[q_1,\ldots,q_n]$ be a component of a level, $n\geq3$.
An \textbf{irregular pair} of $Q^{\emph{c}}$ is defined to be pairwise integers $(i,j),
1\leq i\leq j\!-\!2\leq n\!-\!2$  such that
\begin{align}
              & q_i = q_j  \label{eq-irre1}\\
\textrm{or~~} & q_i \sim q_j, ~ q_i\notin \{q_{j-1}, q_{j+1}\}, ~  q_j\notin \{q_{i-1}, q_{i+1}\} \label{eq-irre2}
\end{align}
where $\{q_{i-1}, q_{i+1}\}$ degenerates into $\{q_{1}\}$ ($\{q_{n}\}$, respectively) when one of the subscripts of $q$ in $\{q_{i-1}, q_{i+1}\}$ is $0$ ($n+1$, respectively). The vertices $q_i,q_j$ are called the end-vertices of the irregular pair $(i,j)$.
\end{definition}

\begin{definition}\label{def-proper}
Let $Q$ be a level of a mesh and let $Q'$ be a sublevel of $Q$. Then $Q'$ is called a \textbf{proper} sublevel of $Q$ if each component of $Q'$ contains no end-vertices of irregular pairs satisfying Equation~(\ref{eq-irre1}) and if each component of $Q'$ contains at most an end-vertex of irregular pairs satisfying Equation~(\ref{eq-irre2}).
\end{definition}
\begin{figure}[bhtp] \centering
\includegraphics[width=0.3\textwidth]{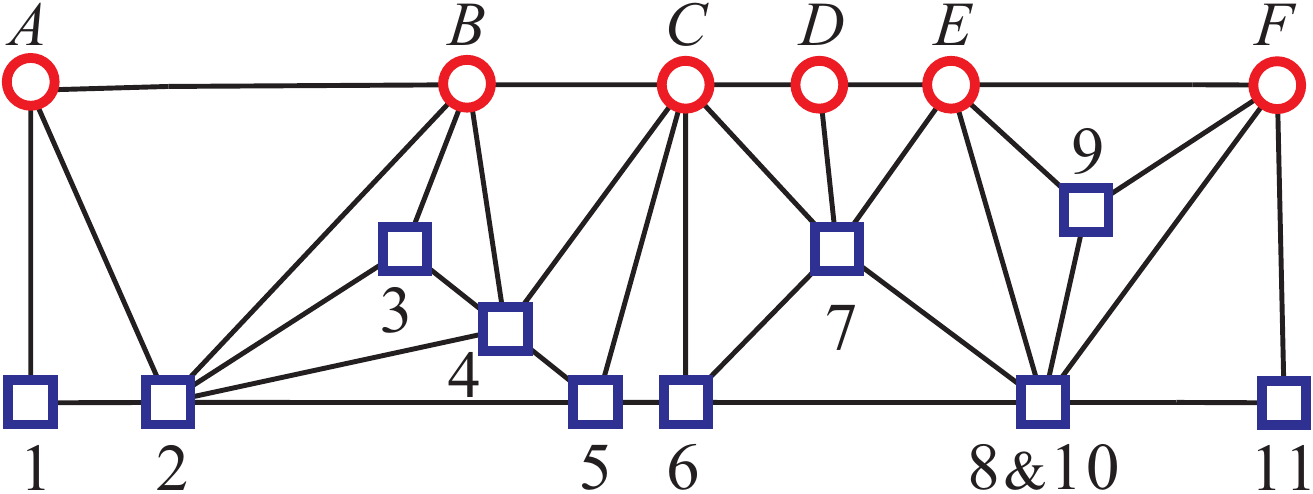}
\caption{An example of irregular pairs and proper sublevels. Let $Q=[1,2,3,4,5,6,7,8,9,10,11]$ be a level which is also a component of itself. Then $(2,4), (2,5), (6,8), (6,10), (8,10)$ are irregular pairs of $Q$, and $[1,2,3,6,7,9]$, $[3,4,5,6,7,9]$ are both proper sublevels of $Q$, but $[1,2,3,4]$, $[3,4,5,6,7,8]$ are not proper sublevels of $Q$ because $(2,4)$ is an irregular pair and because the vertex $8/10$ is an end-vertex of an irregular pair satisfying Equation~(\ref{eq-irre1}), respectively.} \label{fig-proper}
\end{figure}

We show an example of proper sublevels in Fig.~\ref{fig-proper}.
Now we give an intuitive explanation to help readers understand the above definitions.
The proper sublevel is used for determining the level sequence in the \textsf{Stratify} phase.
Suppose that $\mathcal{L}_j,j=1,\ldots,i-1$ are given. We shall choose the elements from the adjacent elements of $\mathcal{L}_{i-1}$. If the adjacent elements of $\mathcal{L}_{i-1}$ are ``well organized" (i.e. they can be arranged to be adjacent one by one, such as the first example of Fig.~\ref{fig-exam2}), then $\mathcal{L}_i$ is simply chosen to be all these elements. However there exist cases in which all these elements are not ``well organized". See the second example of Fig.~\ref{fig-exam2}: whatever combination you choose (e.g. $GFHIJLK$ or others), the vertices $G, J, L$ have no adjacent elements in the forthcoming level as they are isolated by edges within $\mathcal{L}_{i-1}$ and $\mathcal{L}_{i}$. In order to guarantee that no incorrect connectivity be produced in such a case, we can only choose a subset of all these elements together with some elements from $\mathcal{L}_{i-1}$ to form $\mathcal{L}_{i}$. As we notice that such ``isolated" vertices appear only when irregular pairs exist (e.g. $(1,3),(5,7),(5,8)$ are irregular pairs of the candidate level $[FGFHIJK]$), we must choose the subset containing no irregular pairs, which make the proper sublevels a good option.

\begin{definition}\label{def-align1}{\rm (the \textsf{align1} function for the \textbf{slash}-direction connection)~}
Let $\mathcal{L}_1:=[v_1,\ldots,v_m]$, $\mathcal{L}_2:=[w_1,\ldots,w_n]$ be two levels of a mesh, $m,n\in\mathbb{N}$.
The function $(\mathcal{L}'_1, \mathcal{L}'_2)=\mathsf{align1}(\mathcal{L}_1, \mathcal{L}_2)$ returns the following pairwise levels
\[ \begin{array}{l}
\mathcal{L}'_1 = [ \underbrace{v_1,\ldots,v_1}_{\max(1,d_{1})}, ~ \underbrace{v_2,\ldots,v_2}_{\max(1,d_{2})}, ~\cdots, ~\underbrace{v_m,\ldots,v_m}_{\max(1,d_{m})}] \\
\mathcal{L}'_2 = [ \underbrace{w_1,\ldots,w_1}_{\max(1,d'_{1})}, \underbrace{w_2,\ldots,w_2}_{\max(1,d'_{2})}, \cdots, ~\underbrace{w_n,\ldots,w_n}_{\max(1,d'_{n})}]
\end{array} \]
where
\begin{align*}
& d_{1} = |\boldsymbol{N}(v_1)\cap\mathcal{L}_2|,~ d_{i} = |\boldsymbol{N}(v_i)\cap\mathcal{L}_2|-1, ~ i=2,\ldots,m \\
& d'_{n} \!=\! |\boldsymbol{N}(w_n)\cap\mathcal{L}_1|, d'_{i}=\! |\boldsymbol{N}(w_i)\cap\mathcal{L}_1|\!-\!1,~ i\!=\!1,\ldots,n\!-\!1
\end{align*}
\end{definition}

\subsection{The \textsf{isomatrix} phase}\label{sec-42}

\begin{figure*}[t]
\begin{center}
\includegraphics[width=1.0\textwidth]{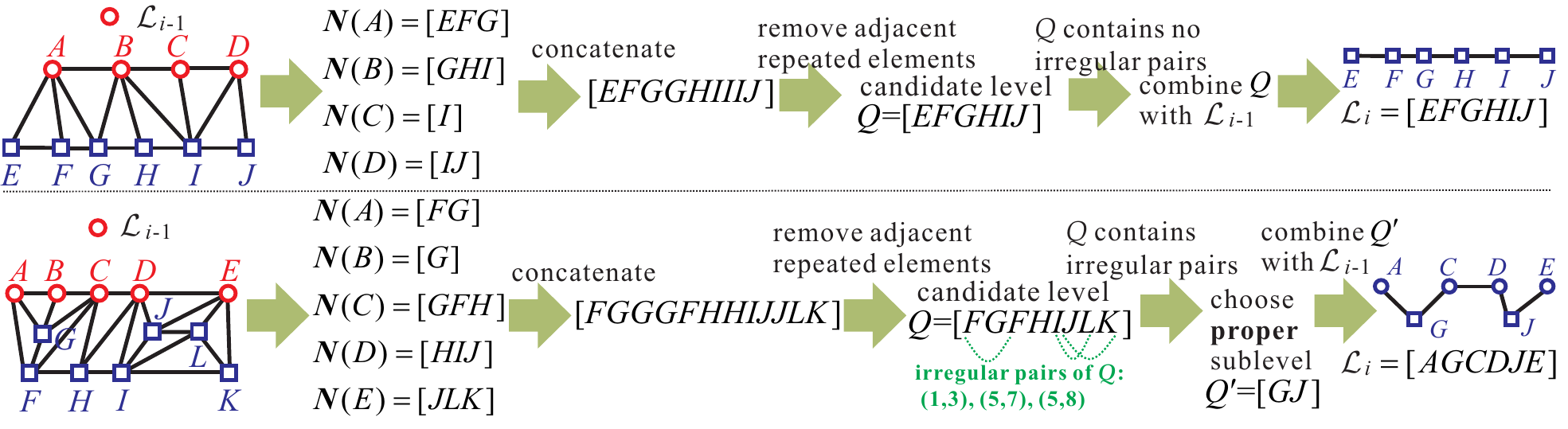}
\caption{Two examples for showing how to determine $\mathcal{L}_i$ based on $\mathcal{L}_{i-1}$ and its neighbors.} \label{fig-exam2}
\end{center}
\end{figure*}

\begin{figure}[h!]
\begin{center}
\includegraphics[width=0.35\textwidth]{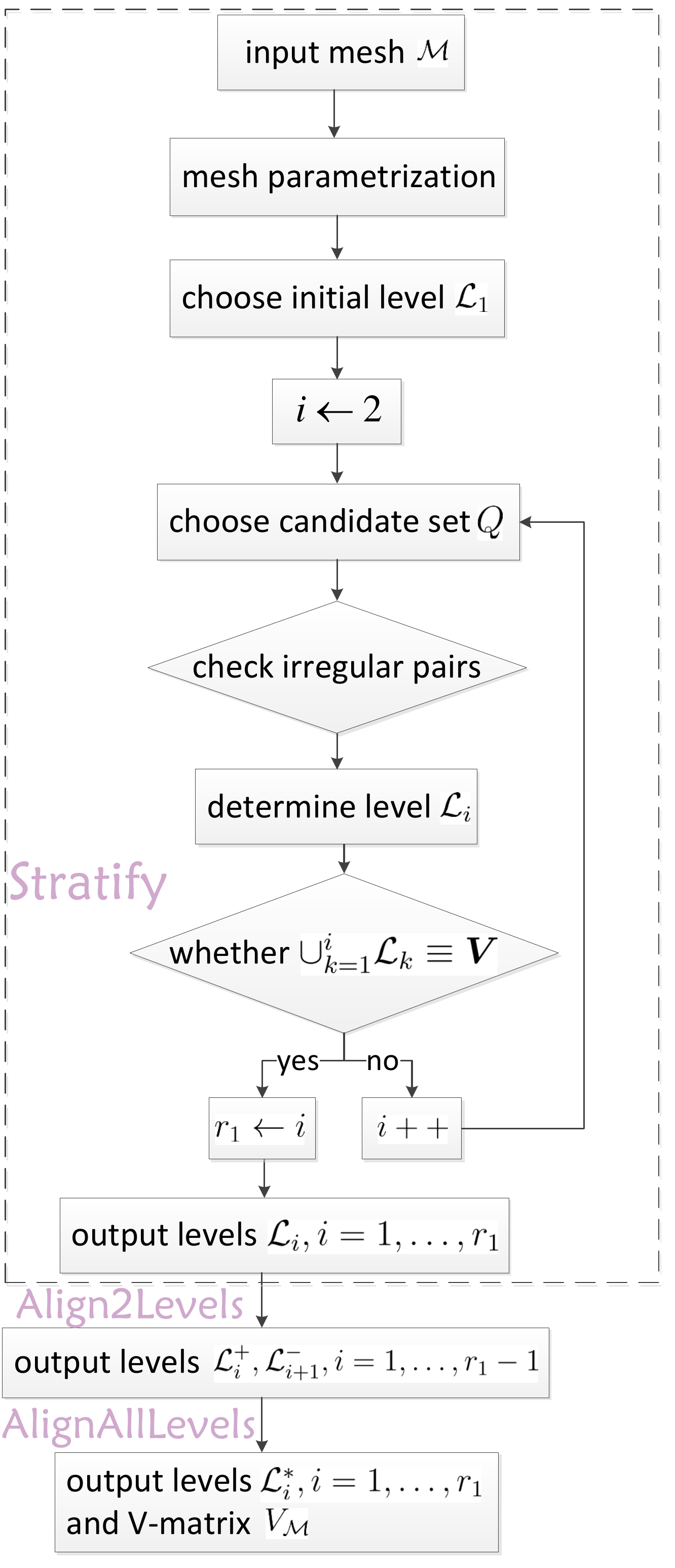}
\caption{An \textsf{isomatrix} flowchart: generating a V-matrix $V_{\mathcal{M}}$ with respect to an open genus-zero triangular mesh $\mathcal{M}$.}\label{fig-flowchart}
\end{center}
\end{figure}

A rough procedure for the \textsf{isomatrix} phase for open genus-zero triangular meshes is described in Fig.~\ref{fig-flowchart}, where the \textsf{stratify} phase
recursively determines the level $\mathcal{L}_i$ by using $\mathcal{L}_{i-1}$ and its neighboring elements, until all vertices are traversed. Before going to the detailed algorithm
of Section~\ref{sec-45}, we illustrate the main idea using two examples of Fig.~\ref{fig-exam2}. When the neighboring elements of $\mathcal{L}_{i-1}$ contain no irregular pairs (see the first example, where the vertices of the component $[EFGHIJ]$ are simply connected one by one), then $\mathcal{L}_{i}$ is simply chosen to be $[EFGHIJ]$; otherwise (see the second example, where $I\sim K$ occurs while $I,K$ are located in nonadjacent positions within $[FGHIJK]$), we choose $\mathcal{L}_{i}$ to be the union of a few vertices of $F,G,H,I,J,K$ together with some vertices of $\mathcal{L}_{i-1}$. We briefly outline the three sub-phases: \textsf{stratify}, \textsf{Align2Levels}, \textsf{AlignAllLevels} as follows, and give the detailed algorithm in the next section.

\begin{window}[12, l,{\mbox{
\includegraphics[width=0.24\textwidth]{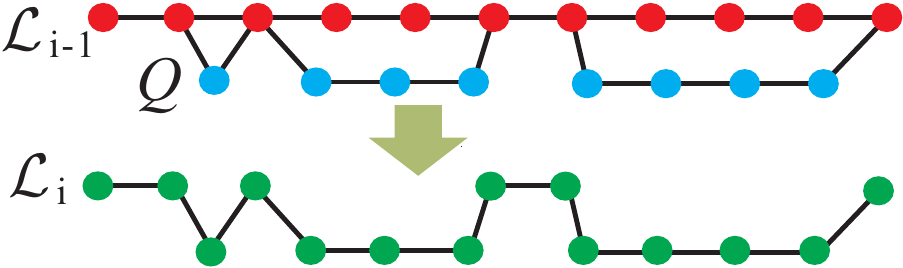}}},{}]
\textbf{\textsf{Stratify}} \quad
The purpose of this phase is to obtain a level sequence $\mathcal{L}_{i}$, $i=1,\ldots,r_1$. We first parametrize an open genus-zero  mesh $\mathcal{M}$ onto the planar domain $[0,1]\times[0,1]$ by using Tutte's parametrization \cite{tuttehow-63}, and choose all the vertices whose ordinate is one to be the initial level $\mathcal{L}_1$.
For $i=2,3,\ldots$, we recursively determine the level $\mathcal{L}_i$ in two steps: (i) obtain a candidate level $Q$ by ordering each $\boldsymbol{N}(\mathcal{L}_{i-1}(j))\setminus\cup_{k=1}^{i-1}\mathcal{L}_k$ in counterclockwise direction with respect to $\mathcal{L}_{i-1}(j)$,
and removing repeated elements within adjacent locations of $Q$; (ii) if $Q$ contains no {irregular pairs}, then we obtain $\mathcal{L}_i$ by \emph{combining} $Q$ \emph{with} $\mathcal{L}_{i-1}$ (see the figure on the left); otherwise, we obtain $\mathcal{L}_i$ by \emph{combining} a proper sublevel of $Q$ \emph{with} $\mathcal{L}_{i-1}$. The whole loop ends when $\cup_{k=1}^i\mathcal{L}_k=\boldsymbol{V}$ holds.
\end{window}

\textbf{\textsf{Align2Levels}}  \quad
The purpose of this phase is to obtain a level sequence $\mathcal{L}_i^+, \mathcal{L}_{i+1}^-, i=1,\ldots,r_1-1$
such that: $ |\mathcal{L}_i^+|=|\mathcal{L}_{i+1}^-|$, and  all the edge-sets
induced by $\mathcal{L}_i^+, \mathcal{L}_{i+1}^-$ are equal to the edge-set of $\mathcal{M}$.
We do it in this way: for each distinct element $v\in\mathcal{L}_i\setminus\mathcal{L}_{i+1}, 1\leq i\leq r_1-1$,
we check the number of its neighbors in $\mathcal{L}_{i+1}$, and add $v$ in $\mathcal{L}_i$ repeatedly to maintain the connectivity, then we repeat the same step by adding $v$
in $\mathcal{L}_{i+1}$ for  $v\in\mathcal{L}_{i+1}\setminus\mathcal{L}_{i}$.

\textbf{\textsf{AlignAllLevels}}\quad
So far, we obtain a level sequence $\mathcal{L}_i^+, \mathcal{L}_{i+1}^-, i=1,\ldots,r_1-1$ such that $|\mathcal{L}_i^+|=|\mathcal{L}_{i+1}^-|$,
where $\mathcal{L}_{i}^-, \mathcal{L}_{i}^+$ contain the same elements with different numbers in them.
We denote $v_1,\ldots,v_m$ to be all different elements of $\mathcal{L}_{i}^-$ in turn.
The purpose of this phase is to obtain another level sequence $\mathcal{L}_i^*, i=1,\ldots,r_1$ such that all the levels $\mathcal{L}_i^*$ have the same length,
and the edge-set induced by $\mathcal{L}_i^+, \mathcal{L}_{i+1}^-$ equals to the edge-set induced by $\mathcal{L}_i^*, \mathcal{L}_{i+1}^*$.
We do it in this way: first for $i=2,\ldots,r_1-1$, we add repeated elements in $\mathcal{L}_i^+$
such that $|\mathcal{L}_i^+|_{v_j}\geq|\mathcal{L}_i^-|_{v_j}, j=1,\ldots,m$, and we simultaneously add repeated elements in $\mathcal{L}_{i+1}^-$ to maintain the same length of $\mathcal{L}_i^+, \mathcal{L}_{i+1}^-$ and the same  edge-set induced by $\mathcal{L}_i^+, \mathcal{L}_{i+1}^-$; then for $i=r_1-1,r_1-2,\ldots,1$, we add repeated elements in $\mathcal{L}_{i+1}^-$
such that $|\mathcal{L}_i^-|_{v_j}\geq|\mathcal{L}_i^+|_{v_j}, j=1,\ldots,m$.
That makes $\mathcal{L}_i^-, \mathcal{L}_i^+$ become exactly the same set, and consequently gives the level $\mathcal{L}_i^*$. Finally, we obtain
a V-matrix $V_{\mathcal{M}}$ with respect to $\mathcal{M}$  by setting $\mathcal{L}_i^*$ as its $i$-th row vector \footnote{The \textsf{isogim} phase generates a CGIM array, by using a linear transform on the three-dimensional matrix whose elements correspond to the coordinates of vertices of the V-matrix $V_{\mathcal{M}}$ obtained in the \textsf{isomatrix} phase. The \textsf{reconstruct} phase accordingly transforms the CGIM array to a vertex-set and an edge-set of the reconstructive mesh. Such two phases are trivial hence we omit the details.}:
\begin{equation} \label{eq-isomatrix}
V_{\mathcal{M}}:=\begin{bmatrix} \mathcal{L}_1^* \\ \mathcal{L}_2^* \\ \vdots \\ \mathcal{L}_{r_1}^*  \end{bmatrix}
\end{equation}

\subsection{CGIM properties}
Let $V_{\mathcal{M}}$ be a V-matrix with respect to a mesh $\mathcal{M}$ with $m$ rows and $n$ columns.
We denote $V_{\mathcal{M}}(i,j)$ to be the element in the $i$-th row, $j$-th column of $V_{\mathcal{M}}$, and denote $V_{i}^{\mathsf{row}}$
to be the level consisting of the $i$-th row of $V_{\mathcal{M}}$, $i=1,\ldots,m$, $j=1,\ldots,n$.
Then $V_{\mathcal{M}}$ is said to be \emph{connectivity-preserving} to $\mathcal{M}$,
if the vertex-set $ \cup_{i=1}^{m}\cup_{j=1}^{n} V_{\mathcal{M}}(i,j)$ induced by $V_{\mathcal{M}}$ equals to the vertex-set of $\mathcal{M}$,
and if the edge-set $\cup_{i=1}^{m-1} \boldsymbol{edge}(V_{i}^{\mathsf{row}}; V_{i+1}^{\mathsf{row}})$ induced by $V_{\mathcal{M}}$ equals to the edge-set of $\mathcal{M}$.

We can check that the V-matrix in the second column of both examples of Fig.~\ref{fig-igim1} is connectivity-preserving to the original mesh respectively.
The following property shows estimates of reconstruction errors of CGIMs.

\vspace{2mm}\noindent\textbf{Property~}
Let $V_{\mathcal{M}}$ be a V-matrix with respect to $\mathcal{M}$ and let $\mathcal{M}'$  be the reconstructed mesh generated from $V_{\mathcal{M}}$ by using a $b$-bit CGIM.
Suppose that $V_{\mathcal{M}}$ is connectivity-preserving to $\mathcal{M}$.
Then the max Hausdorff distance and the root-mean-square Hausdorff distance between $\mathcal{M}'$ and $\mathcal{M}$  are both bounded by
\[ {\sqrt{3}(\max value(\mathcal{M})-\min value(\mathcal{M}))}/(2(2^b-1)) \]

\subsection{An algorithm for the \textsf{isomatrix} phase}\label{sec-45}

This section gives a complete algorithm for the \textsf{isomatrix} phase
which produces a V-matrix preserving the connectivity of the original mesh. Such a property is proved in Section~A of the supplementary material.

\begin{enumerate}
\item[] \textbf{input}: an open genus-zero triangular mesh $\mathcal{M}$.

\item[]  \textbf{output}: the \emph{row resolution} $r_1$, the \emph{column resolution} $r_2$, a level sequence $\mathcal{L}_{i}^{*}$, $i=1,\ldots,r_1$ and a V-matrix $V_{\mathcal{M}}$.

\renewcommand{\labelenumi}{(\arabic{enumi})}
\item\label{step-para}  parametrize $\mathcal{M}$ onto the planar domain $[0,1]\times[0,1]$ using Tutte's parametrization \cite{tuttehow-63}.

\item\label{step-l1}  set $i\leftarrow 2$, and choose the initial level $\mathcal{L}_1$ to be all vertices whose ordinate is one over the parametrized domain, such that
\[ 0=x(\mathcal{L}_1(1))<x(\mathcal{L}_1(2))<\cdots<x(\mathcal{L}_1({\rm end}))=1 \]
where $x(\cdot)$ denotes the abscissa of a vertex in $\mathbb{R}^2$.

\item\label{step-order}  for $ j=1, \ldots, |\mathcal{L}_{i-1}|$, order  $Q_j\!:=\!\boldsymbol{N}(v_j)\setminus\cup_{k=\!1}^{i-1}\mathcal{L}_k$ by
\[ ~~\angle v_{j-\!1}v_jQ_j(1)  \!<\! \angle v_{j-\!1}v_jQ_j(2) \!<\! \cdots \!<\! \angle v_{j-\!1}v_jQ_j({\rm end}) \]
where $v_j:=\mathcal{L}_{i-1}(j)$,
 $\overrightarrow{v_1v_0}:=\overrightarrow{v_1}$, and  $\angle a_1a_2a_3\in[0,2\pi)$ denotes the angle from a planar vector $\overrightarrow{a_2a_1}$
to another planar vector $\overrightarrow{a_2a_3}$ in counterclockwise direction.

\item\label{step-setq1}  set $Q\leftarrow\langle Q_1, Q_2,\ldots, Q_{|\mathcal{L}_{i-1}|}\rangle$, and remove $Q(k+1)$ which satisfies $Q(k)=Q(k+1)$, $1\leq k\leq |Q|-1$.

\item\label{step-component1}  compute all the components $Q_1^{\textrm{c}}, \ldots, Q_m^{\textrm{c}}$ of $Q$.

\item \label{step-proper}
for each component $Q_j^{\textrm{c}}$ with irregular pairs, compute a proper sublevel $Q'$ of $Q_j^{\textrm{c}}$, $1\leq j\leq m$, and set $Q_j^{\textrm{c}}\leftarrow Q'$.

\item\label{step-setq2}  $Q\leftarrow\langle Q_1^{\textrm{c}}, \ldots, Q_m^{\textrm{c}}\rangle$.

\item\label{step-component2}  compute all the components $Q_1^{\textrm{c}}, \ldots, Q_m^{\textrm{c}}$ of $Q$.

\item\label{step-computkj}  for each component $Q_j^{\textrm{c}}$ of $Q$,
compute the smallest index  of $Q_j^{\textrm{c}}(1)$'s adjacent elements in $\mathcal{L}_{i-1}$
and the greatest index of $Q_j^{\textrm{c}}({\rm end})$'s adjacent elements in $\mathcal{L}_{i-1}$, i.e.
     \begin{eqnarray*}
  k_j^{-} &=& \min\{1\leq k\leq |\mathcal{L}_{i-1}|: Q_j^{\textrm{c}}(1)\sim\mathcal{L}_{i-1}(k)\}, \\
  k_j^{+} &=& \max\{1\leq k \leq |\mathcal{L}_{i-1}|: Q_j^{\textrm{c}}({\rm end})\sim\mathcal{L}_{i-1}(k)\}, \\
 j~~&=&1,\ldots,m
     \end{eqnarray*}

\item  \label{step-deterLi}
determine the level $\mathcal{L}_i$:
\[    \mathcal{L}_i \leftarrow \langle\mathcal{L}_{i-1}(1:k_1^{-}), Q_1^{\textrm{c}}, \cdots, Q_m^{\textrm{c}}, \mathcal{L}_{i-1}(k_m^{+}:{\rm end})\rangle \]

\item \label{step-align1}  by using Definition~\ref{def-align1}, compute
        \[(P'_j, Q'_j)=\mathsf{align1}(\mathcal{L}_{i-1}(k_j^{-}:k_j^{+}), Q_j^{\textrm{c}}), ~~~ j=1,\ldots,m.\]

\item \label{step-deterLi+}  determine the levels $\mathcal{L}_{i-1}^{+}$, $\mathcal{L}_{i}^{-}$:
        \begin{align*}
&        \mathcal{L}_{i-\!1}^{+} \leftarrow \langle\mathcal{L}_{i-1}(1: k_1^{-}), P'_1, \cdots, P'_m, \mathcal{L}_{i-1}(k_m^{+}:{\rm end}) \rangle  \\
&        \mathcal{L}_{i}^{-} ~\leftarrow \langle\mathcal{L}_{i-1}(1: k_1^{-}), Q'_1, \cdots, Q'_m, \mathcal{L}_{i-1}(k_m^{+}:{\rm end}) \rangle
        \end{align*}

\item \label{step-deterLi2} if $Q(1)\!\sim\!\mathcal{L}_{i-1}(1)$, $Q(1)\in\boldsymbol{V_B}$,
 then  remove the first term $\mathcal{L}_{i-1}(1\!:\! k_1^{-})$ from $\mathcal{L}_i$, $\mathcal{L}_{i-1}^{+}$ and $\mathcal{L}_{i}^{-}$.
 If $Q({\rm end})\!\sim\!\mathcal{L}_{i-1}({\rm end})$, $Q({\rm end})\in\boldsymbol{V_B}$,
 then  remove the last term $\mathcal{L}_{i-1}(k_m^{+}\!:\!{\rm end})$ from $\mathcal{L}_i$, $\mathcal{L}_{i-1}^{+}$ and $\mathcal{L}_{i}^{-}$.
\footnote{The reason we do Step~(\ref{step-deterLi2}) is that if $Q(1)\sim\mathcal{L}_{i-1}(1)$, $Q(1)\in\boldsymbol{V_B}$, then $Q(1)$ is capable of being the first element of $\mathcal{L}_i$ and hence the level $\mathcal{L}_{i-1}(1:k_1^-)$ is unnecessary for being a sublevel of $\mathcal{L}_i$, $\mathcal{L}_{i-1}^{+}$, $\mathcal{L}_{i}^{-}$; the reason is similar for $Q({\rm end})$.}

\item \label{step-endloop}  if $\cup_{j=1}^{i}\mathcal{L}_j=\boldsymbol{V}$ holds, then set $r_1\leftarrow i$ and
goto Step~(\ref{step-fori2}); otherwise, $i++$, and goto Step~(\ref{step-order}).

\item \label{step-fori2} for $i=2,\ldots, r_1-1$,
      $(\mathcal{L}_{i}^{+}, \mathcal{L}_{i+1}^{-}) \!\leftarrow\! \mathsf{align2}(\mathcal{L}_{i}^{-}, \mathcal{L}_{i}^{+}, \mathcal{L}_{i+1}^{-})$.

\item \label{step-computr2}  $\mathcal{L}_{r_1-1}^{*}\leftarrow\mathcal{L}_{r_1-1}^{+}$, ~~ $\mathcal{L}_{r_1}^{*}\leftarrow\mathcal{L}_{r_1}^{-}$,~~$r_2\leftarrow|\mathcal{L}_{r_1}^{*}|$.

\item \label{step-forir1}
for $i=r_1-2,\ldots, 1$, ~$(\sim, \mathcal{L}_{i}^{*})\leftarrow\mathsf{align2}(\mathcal{L}_{i+1}^{+}, \mathcal{L}_{i+1}^{-}, \mathcal{L}_{i}^{+})$, where
the first output variable is denoted by $\sim$ as it is not used.  

\item \label{step-endvmatrix}  obtain a V-matrix $V_{\mathcal{M}}$ using Equation~(\ref{eq-isomatrix}).
\end{enumerate}
\begin{algorithm}[thbp]
\caption{the \textsf{align2} function} \label{alg-align2}
\SetCommentSty{emph}
\SetKwFunction{aligntwo}{\textsf{align2}}
\SetKwInOut{Input}{input}\SetKwInOut{Output}{output}
\Input{three levels $\mathcal{L}_1$, $\mathcal{L}_2$, $\mathcal{L}_3$}
\Output{two levels $\mathcal{L}_2$, $\mathcal{L}_3$}
denote $v_i, i=1,\ldots,t$ to be all different elements of $\mathcal{L}_1$ in turn, where $t$ is the number of different elements in $\mathcal{L}_1$\;
\For{~$i=1,2,\ldots,t$~}
    {
      $d\leftarrow |\mathcal{L}_1|_{v_i}- |\mathcal{L}_2|_{v_i}$\;
     \If{~~$d\geq1$~\nllabel{alg-ifd1}}
        {
          compute the smallest index $a_1$ and the largest index $a_2$ of $v_i$ in $\mathcal{L}_2$, respectively\;

          denote $w_1,\ldots,w_k$ to be all \emph{different} elements of $\mathcal{L}_3(a_1:a_2)$ in turn\;

          insert $d$ number of $v_i$ into $\mathcal{L}_2$ so that all $v_i$'s are in adjacent locations\; \nllabel{alg-insert1}

          $j\leftarrow 1$\;

          \While{~$d\geq1$~}
              {
              insert a $w_j$ in $\mathcal{L}_3$ so that all $w_j$'s are in adjacent locations\; \nllabel{alg-insert2}
              $d--$\;
              $j\leftarrow (j\!\mod k)+1$\;
              }
        }
    }
\end{algorithm}

\begin{remark}\label{rem-igim1}
Such a CGIM algorithm is parametrization-free. This is because the following inequality $$\angle \mathcal{L}_{i-1}(j-1)\mathcal{L}_{i-1}(j)q \neq \angle \mathcal{L}_{i-1}(j-1)\mathcal{L}_{i-1}(j)q'$$
holds for any two distinct adjacent vertices $q$, $q'$ of $Q_j$, $j=1,\ldots,|\mathcal{L}_{i-1}|$, and such an inequality  depends not on the parametrization but on the connectivity of $\mathcal{M}$.
\end{remark}

\begin{remark}\label{rem-igim3}
In Algorithm~\ref{alg-align2} line~\ref{alg-insert1}, we insert all $v_i$'s into $\mathcal{L}_2$ so that all $v_i$'s are in adjacent locations in the new $\mathcal{L}_2$;
so do the $w_j$'s in line~\ref{alg-insert2}. This is guaranteed because all $v_i$'s are in adjacent locations of $\mathcal{L}_{i}^{+}$, as well as in adjacent locations of $\mathcal{L}_{i+1}^{-}$, according to Lemma~4 of Section~A of the supplementary material.
\end{remark}

\subsection{A guide of the \textsf{isomatrix} phase}
\begin{figure*}[thbp]
\begin{center}
\includegraphics[width=1.0\textwidth]{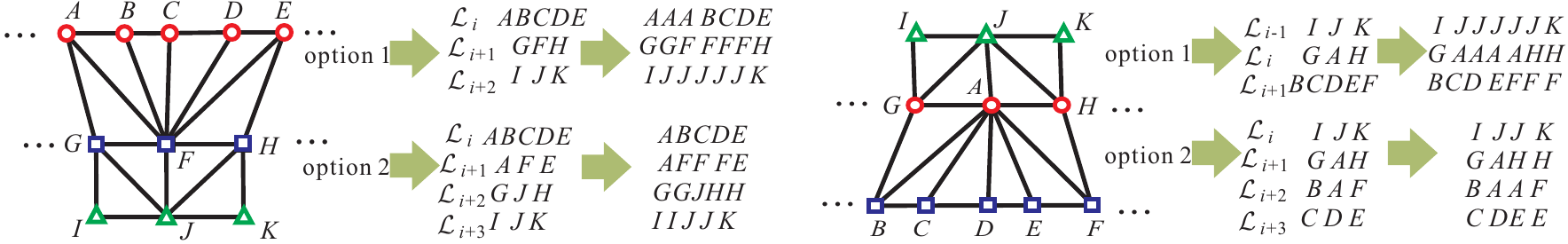}
\caption{Different vertex groupings produce V-matrices with different resolutions.} \label{fig-columnreso}
\end{center}
\end{figure*}

We repeat the \textsf{isomatrix} phase (Steps (1)-(\ref{step-endvmatrix})) using the second example of Fig.~\ref{fig-exam2} in this section.  Steps~(1),~(\ref{step-l1}) choose the initial level $\mathcal{L}_1=[{ABCDE}]$. Step~(\ref{step-order}) arranges all the neighbors of $\mathcal{L}_1$ according to the angles with respect to the associated adjacent elements in $\mathcal{L}_1$, which gives
\begin{align*} & \boldsymbol{N}({A})=[{FG}],~~~~~\boldsymbol{N}({B})=[{G}],~~~~\boldsymbol{N}({C})=[{GFH}],\\
& \boldsymbol{N}({D})=[{HIJ}], ~~~\boldsymbol{N}({E}) = [{JK}]
\end{align*}
Step~(\ref{step-setq1}) concatenates the above levels together and removes the repeated elements that are located in adjacent positions, which give a candidate level
\[Q= [{FGFHIJK}]\]
Step~(\ref{step-component1}) computes all the components of $Q$, which are $Q$ itself merely, denoted by $Q_1^{\textrm{c}}$.
Step~(\ref{step-proper}) finds irregular pairs $(1,3),(5,7)$ of $Q_1^{\textrm{c}}$ and computes proper sublevels of $Q_1^{\textrm{c}}$ which are $[{G}]$ and $[{J}]$ \footnote{The choice of proper sublevels is not unique. We can also choose $[{G}], [{HIJ}]$  as proper sublevels of $Q$. Because the choice of proper sublevels is related to the size of the V-matrix we obtain, we shall give a modified \textsf{isomatrix} algorithm for generating a CGIM array of smaller size by careful choice of proper sublevels in Algorithm~\ref{alg-modifyiso}.}.
Steps~(\ref{step-setq2}),~(\ref{step-component2}) concatenate all sublevels together to form a level denoted by $Q=[{GJ}]$ and recompute all components of $Q$, which are
\[ Q_1^{\textrm{c}}=[{G}], \quad Q_2^{\textrm{c}} =[{J}]. \]
Steps~(\ref{step-computkj}),~(\ref{step-deterLi}) determine $\mathcal{L}_2$ by first combining each component $Q_j^{\textrm{c}},j=1,2$ with $\mathcal{L}_{1}$ using the first adjacent element of $Q_j^{\textrm{c}}(1)$ in $\mathcal{L}_{1}$ and last adjacent element of $Q_j^{\textrm{c}}(\mbox{end})$ in $\mathcal{L}_{1}$, and then concatenating them together, which gives  \[\mathcal{L}_2=[{AGCDJE}]\]
where $A, C$ are the first and the last adjacent elements of $[{G}]$, and $D, E$ are the first and the last adjacent elements of $[{J}]$. Steps~(\ref{step-align1})-(\ref{step-deterLi2}) constitute the \textsf{Align2Levels} phase, which first
gives \[ P'_1=[{ABC}], ~Q'_1=[{AGC}]; ~~~  P'_2=[{DDE}], ~Q'_2=[{DJE}]\] and then yields
\[ \mathcal{L}_1^+=[{ABCDDE}] \quad  \mathcal{L}_2^-=[{AGCDJE}]\]
Step~(\ref{step-endloop}) repeats the above procedure for $i=3$, finds irregular pairs ($I\sim K$ holds within the candidate level $[{FHILK}]$) and hence obtain
\begin{align*} &\mathcal{L}_3=[{AGCDJLE}] \\
& \mathcal{L}_2^+ =[{AGCDJJE}]~~~  \mathcal{L}_3^-=[{AGCDJLE}]
\end{align*}
Step~(\ref{step-endloop}) repeats the above procedure for $i=4$, finds no irregular pairs and hence obtain
\[ \mathcal{L}_4\!=\![{FHIK}]~~~~  \mathcal{L}_3^+ \!=\![{AGCDJLE}]~~~~  \mathcal{L}_4^-\!=\![{FFHIIKK}]\]
So far all the vertices are traversed, then we set $r_1=4$ and go to Step~(\ref{step-fori2}) to update $\mathcal{L}_2^+,\mathcal{L}_3^-,\mathcal{L}_3^+,\mathcal{L}_4^-$.
However as in this case $|\mathcal{L}_i^-|_{v}\leq |\mathcal{L}_i^+|_{v}$ holds for all $v\in\mathcal{L}_i,~i=2,3$, i.e. the condition in line~\ref{alg-ifd1} of Algorithm~\ref{alg-align2} is always invalid, those four level sequences remain the same. Then Step~(\ref{step-computr2}) obtains
\[ \mathcal{L}_3^*=[{AGCDJLE}]\quad \mathcal{L}_4^*=[{FFHIIKK}]\]
and set $r_2=7$. In Step~(\ref{step-forir1}), because $\mathcal{L}_2^+$ contains one more $J$'s than $\mathcal{L}_2^-$, we insert a corresponding element $D$ in $\mathcal{L}_1^+$ and give
\[ \mathcal{L}_1^*=[{ABCDDDE}]\]
Finally, Step~(\ref{step-endvmatrix}) gives the V-matrix
\[\begin{pmatrix}
{A} & {B} & {C} & {D} & {D} & {D} & {E} \\
{A} & {G} & {C} & {D} & {J} & {J} & {E}\\
{A} & {G} & {C} & {D} & {J} & {L} & {E}\\
{F} & {F} & {H} & {I} & {I} & {K} & {K} \end{pmatrix}\]
which can be verified to be connectivity-preserving to the second example of Fig.~\ref{fig-exam2}.

\section{Other algorithms for CGIMs}\label{sec-5}

This section gives algorithms for CGIMs of smaller resolutions in Section~\ref{sec-51}, and for reconstructing meshes from compressed CGIMs in Section~\ref{sec-52}. We introduce the idea of CGIMs for treating closed genus-zero meshes in Section~B of the supplementary material.

\subsection{CGIMs with smaller resolutions}\label{sec-51}

Compared with traditional GIMs, the greatest disadvantage of CGIMs is that they have blow-up resolutions in order to entirely preserve the connectivity of a mesh. Such a disadvantage decreases the efficiency of mesh compression using CGIMs.
We shall propose a modified \textsf{isomatrix} phase for CGIMs of smaller resolutions (Algorithm~\ref{alg-modifyiso}). Compared with the \textsf{isomatrix} phase in Section~\ref{sec-45}, the modification consists of two aspects. One aspect comes from the observation that the column resolution of CGIMs is much greater than the row resolution (varying from one and a half times to two and a half times in our experiments), which drives us adjust the \textsf{stratify} phase to re-group the vertex-set in order to produce a smaller column resolution; the other is to check each row and each column of the V-matrix $V_{\mathcal{M}}$ to see
whether any row or any column of $V_{\mathcal{M}}$ is removable (Algorithm~\ref{alg-modifyiso}, lines~\ref{alg-removrow}-\ref{alg-removcol}), i.e. whether
\begin{align}
&\hspace{-2mm} \big(\!\cup_{2\leq k\leq r_1\atop k\neq i,i+1}\!\!\boldsymbol{edge}(M_{k-\!1}^{\mathsf{row}};\! M_{k}^{\mathsf{row}}) \!\big)\!\cup\! \boldsymbol{edge}(M_{i-\!1}^{\mathsf{row}};\! M_{i+\!1}^{\mathsf{row}}) \!=\!\boldsymbol{E} \label{eq-remov1}\\
&\hspace{-2mm} \big(\!\cup_{2\leq k\leq r_2\atop k\neq j,j+1}\!\!\boldsymbol{edge}(M_{k-\!1}^{\mathsf{col}};\! M_{k}^{\mathsf{col}}) \!\big)\!\cup\! \boldsymbol{edge}(M_{j-\!1}^{\mathsf{col}};\! M_{j+\!1}^{\mathsf{col}}) \!=\!\boldsymbol{E}  \label{eq-remov2}
\end{align}
hold for some $i, j$ respectively, where $M_{i}^{\mathsf{row}}, M_{j}^{\mathsf{col}}$ denote the level consisting of the $i$-th row, the $j$-th column of $V_{\mathcal{M}}$ respectively, and where the set on the left-hand-side represents the edge-set induced by the V-matrix removing the $i$-th row, or removing the $j$-th column of $V_{\mathcal{M}}$ respectively.

Let us explain the first aspect of the modification using Fig.~\ref{fig-columnreso}. In the first example, the CGIM algorithm gives $\mathcal{L}_i=[{ABCDE}]$, $\mathcal{L}_{i+1}=[{GFH}]$, $\mathcal{L}_{i+2}=[{IJK}]$,
and gives a V-matrix of resolutions $3\times7$ in the right hand side (option~1). The column resolution of the V-matrix is large because vertex F has five neighbors in $\mathcal{L}_i$.
However if we choose $\mathcal{L}_{i+1}=[{AFE}]$, $\mathcal{L}_{i+2}=[{GJH}]$, $\mathcal{L}_{i+3}=[{IJK}]$,
then we give a V-matrix of smaller size $4\times5$ (option~2). In the second example, option~1 gives $\mathcal{L}_{i-1}=[{IJK}]$, $\mathcal{L}_{i}=[{GAH}]$, $\mathcal{L}_{i+1}=[{BCDEF}]$,
and gives a V-matrix of resolutions $3\times7$, while option~2 gives $\mathcal{L}_{i-1}=[{IJK}]$, $\mathcal{L}_{i}=[{GAH}]$, $\mathcal{L}_{i+1}=[{BAF}]$, $\mathcal{L}_{i+2}=[{CDE}]$,
and gives a V-matrix of smaller size $4\times4$. The trick is to combine fewer vertices in the next level when the next level or current level contains vertices with too many neighbors in other levels, so that the column resolution of the V-matrix decreases.

\begin{figure*}[thbp]
\begin{center}
\includegraphics[width=1.01\textwidth]{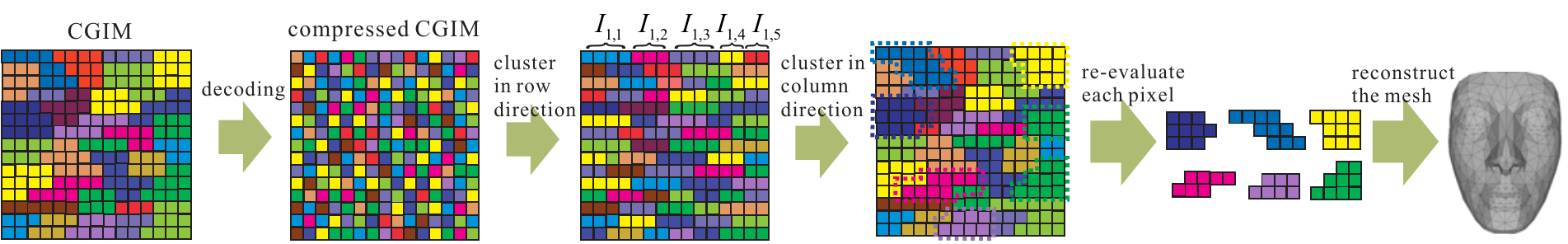}
\caption{ The \textbf{\textsf{cluster}} phase: after a CGIM array is compressed (column two),
we cluster its pixels in rows to obtain a category sequence $I_{i,t}, i=1,\ldots,r_1, t=1,\ldots, x_{i}$ according to the color distance metrics $d_{\mathsf{row}}(i,j)$  (column three), combine $I_{i+1,t}$ with some $I_{i,k}, 1\leq k\leq x_{i}$ according to the category distance metrics $D(I_{i,k},  I_{i+1,t})$  (column four),
re-evaluate each pixel using the mean value of pixels from the same category  (column five),
and finally reconstruct the mesh (column six).} \label{fig-cluster}
\end{center}
\end{figure*}

\begin{algorithm*}[htbp]
\caption{the modified \textsf{isomatrix} algorithm}\label{alg-modifyiso}
\SetCommentSty{emph}
\SetKwFunction{aligntwo}{\textsf{align2}}
\SetKwFunction{execute}{\textbf{execute}}
\SetKwFunction{breakone}{\textbf{break}}
\SetKwInOut{Input}{input}
\SetKwInOut{Output}{output}
\Input{an open genus-zero mesh $\mathcal{M}$, and the threshold $\alpha$}
\Output{a V-matrix $V_{\mathcal{M}}$ which is connectivity-preserving to $\mathcal{M}$}
parametrize $\mathcal{M}$ onto the planar domain $[0,1]\times[0,1]$\;
choose the initial level $\mathcal{L}_1$ using Step~(\ref{step-l1})\;

       \For{~$i=2,3,\ldots$~}
             {
               order the set $Q_j:=\boldsymbol{N}(\mathcal{L}_{i-1}(j))\setminus\cup_{k=1}^{i-1}\mathcal{L}_k$ using Step~(\ref{step-order}), $j=1,2,\ldots,|\mathcal{L}_{i-1}|$ \;
     set $Q\leftarrow \langle Q_1,\ldots,Q_{|\mathcal{L}_{i-1}|}\rangle$, and remove $Q(k+1)$ which satisfies $Q(k)=Q(k+1)$, $1\leq k\leq|Q|-1$ \;
     compute all the components of $Q$ and denote them to be $Q_1^{\textrm{c}},\ldots,Q_m^{\textrm{c}}$ in turn \;
               \ForEach{~$Q_j^{\textrm{c}}$ which contains irregular pairs, $1\leq j\leq m$~}
                        {
               compute a proper sublevel $Q'$ of $Q_j^{\textrm{c}}$ and set ${Q}_j^{\textrm{c}}\leftarrow Q'$ \nllabel{alg-proper}\;
                        }
               \ForEach{~$Q_j^{\textrm{c}}$ which contains no irregular pairs, $1\leq j\leq m$~\nllabel{alg-foreach}}
                        {
$Q'\leftarrow\{v\in Q_j^{\textrm{c}}: |\boldsymbol{N}(v)\cap\mathcal{L}_{i-1}|\geq\alpha \}$
\tcc*[r]{if $v\in Q_j^{\textrm{c}}$ has many neighbors in $\mathcal{L}_{i-1}$ (such as vertex F in the first example of Fig.~\ref{fig-columnreso}), we only combine $v$ with $\mathcal{L}_{i-1}$}
$Q'' \leftarrow \{w\in Q_j^{\textrm{c}}:  w \mbox{~is the first or last neighbors of~} v \mbox{~in~} Q_j^{\textrm{c}} \mbox{~such that~} v\in\mathcal{L}_{i-1} ~{\rm \&\&}~ |\boldsymbol{N}(v)\cap Q_j^{\textrm{c}}|\geq\alpha\}$
                             \tcc*[r]{if $v\in\mathcal{L}_{i-1}$ has many neighbors in $Q_j^{\textrm{c}}$ (such as vertex A in the second example of Fig.~\ref{fig-columnreso}),  we only combine $v$'s first and last neighbors in $Q_j^{\textrm{c}}$ with $\mathcal{L}_{i-1}$}
                             \If{~$Q'\cup Q''\not\equiv \varnothing$~}
                                {
                                 $Q_j^{\textrm{c}}\leftarrow \langle Q', Q''\rangle$\;
                                }
                        }\nllabel{alg-foreachend}
                         set $Q\leftarrow\langle {Q}_1^{\textrm{c}},\ldots,{Q}_m^{\textrm{c}}\rangle$\;
     compute the levels $\mathcal{L}_i$, $\mathcal{L}_{i-1}^+$, $\mathcal{L}_{i}^-$ based on $Q$ using Steps~(\ref{step-computkj})-(\ref{step-deterLi2})\;
     \If{~$\cup_{j=1}^i \mathcal{L}_j\equiv \boldsymbol{V}$~}{$r_1\leftarrow i$, ~ \breakone \;}
             }

\For{~$i=2,3,\ldots, r_1-1$~}
     {\nllabel{a4-for2}
      $(\mathcal{L}_{i}^{+}, \mathcal{L}_{i+1}^{-}) \leftarrow\aligntwo(\mathcal{L}_{i}^{-}, \mathcal{L}_{i}^{+}, \mathcal{L}_{i+1}^{-})$\;
      \nllabel{alg-align21}
     }
$\mathcal{L}_{r_1-1}^{*}\leftarrow\mathcal{L}_{r_1-1}^{+}$, ~~ $\mathcal{L}_{r_1}^{*}\leftarrow\mathcal{L}_{r_1}^{-}$\;
\For{~$i=r_1-2,r_1-3,\ldots, 2,1$~~}
     {
      $(\sim, \mathcal{L}_{i}^{*})\leftarrow\aligntwo(\mathcal{L}_{i+1}^{+}, \mathcal{L}_{i+1}^{-}, \mathcal{L}_{i}^{+})$\;
      \nllabel{alg-align22}
     }
obtain the V-matrix $V_{\mathcal{M}}$  using Equation~(\ref{eq-isomatrix})\;
\For{~$1\leq i\leq$ the number of rows  of $V_{\mathcal{M}}$~\nllabel{alg-removrow}}
      {
      \If{~Equation~(\ref{eq-remov1}) holds~}
         {
          remove the $i$-th row of $V_{\mathcal{M}}$\;
         }
      }
\For{~$1\leq j\leq$ the number of columns of $V_{\mathcal{M}}$~}
      {
      \If{~Equation~(\ref{eq-remov2}) holds~}
         {
          remove the $j$-th column of $V_{\mathcal{M}}$\;
         }
      }\nllabel{alg-removcol}
\end{algorithm*}
Based on this observation, we modify the \textsf{isomatrix} phase by selecting a smaller number of vertices to combine with the last level for obtaining $Q$ in Steps~(\ref{step-proper}), (\ref{step-setq2}), which increases the row resolution and decreases the column resolution.
Such an operation is given in lines~\ref{alg-foreach}-\ref{alg-foreachend} of Algorithm~\ref{alg-modifyiso},
where $\alpha$ is the threshold for controlling the extent for the novel grouping, i.e. when $\alpha$ is small,
fewer new vertices are added in the next level (which increases the row resolution $r_1$ and accordingly decreases the column resolution $r_2$), and vice versa.

\subsection{Reconstruction from compressed CGIMs}\label{sec-52}
CGIMs consist of pixels, many of which share the same RGB values in order to maintain the connectivity of the mesh,
whereas compressed CGIMs do not have such a property because of quantization errors over each pixel during lossy compression.
Thus, to reconstruct meshes from compressed CGIMs, we add a \textsf{cluster} phase so that the pixels with similar values
are re-evaluated with a common value, chosen to be the average of values of those pixels which are clustered in the same category.
\begin{figure*}[htbp] \centering
\includegraphics[width=1.02\textwidth]{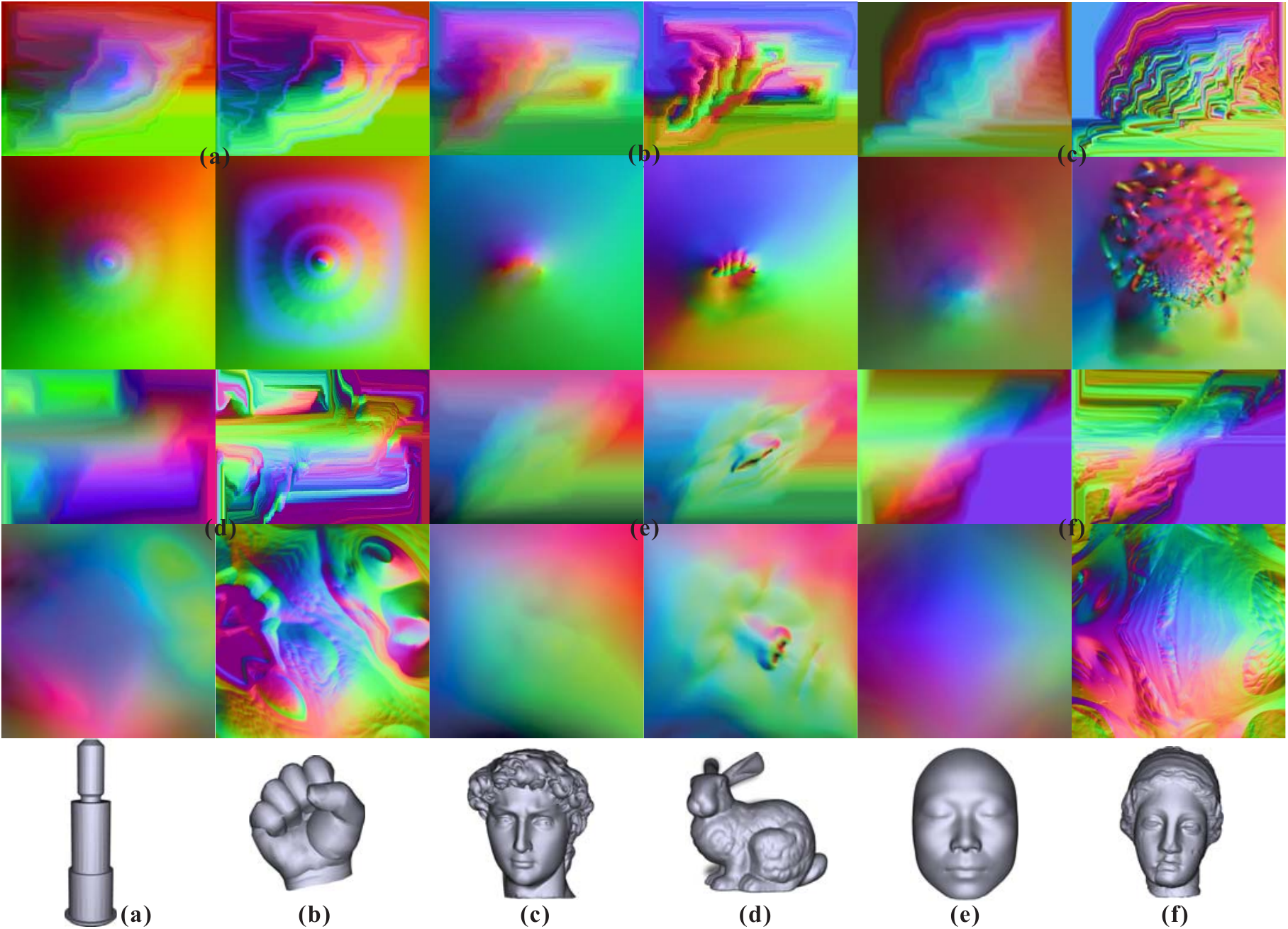}
\caption{CGIMs/CNIMs vs traditional GIMs/NIMs:
(a) $cylinder$, (b) $fist$, (c)  $david$, (d) $bunny$ (e)  $face$, (f) $venus$.
Left-up: CGIMs, right-up: CNIMs; left-down: traditional GIMs, right-down: traditional NIMs.}  \label{fig-igim2}
\end{figure*}

Let $M'\in\mathbb{Z}^{r_1\times r_2\times3}$ be a compressed CGIM array. We record $x_{i}+1,  y_{j}+1$ as the number of different elements in the $i$-th row, the $j$-th column of the V-matrix respectively (such numbers are the additional information we record during CGIM-based lossy compression of meshes), $i=1,\ldots,r_1, j=1,\ldots,r_2$.
Then we compute the color distance metrics $d_{\mathsf{row}}(i,j)$ between pixels $(i,j)$ and $(i,j+1)$, $i=1,\ldots,r_1$, $j=1,\ldots,r_2-1$, and the color distance metrics $d_{\mathsf{col}}(i,j)$ between pixels $(i,j)$ and $(i+1,j)$, $i=1,\ldots,r_1-1, j=1,\ldots,r_2$.
For each $1\leq i\leq r_1$, we put all the pixels of the $i$-th row into the category sequence $I_{i,k}$, $k=1,\ldots,x_i+1$ as follows:
\[   \underbrace{(i,1),\!\cdots\!,(i,j_1)}_{I_{i,1}} \underbrace{(i,j_1\!+\!1),\!\cdots\!,(i,j_2)}_{I_{i,2}} \cdots
 \underbrace{(i,j_{x_i}\!+\!1),\!\cdots\!,(i,r_2)}_{I_{i,x_i+1}}  \]
where $j_1<j_2<\cdots<j_{x_i}$ are chosen to be the indices such that $d_{\mathsf{row}}(i,j_k)$, $k=1,2,\ldots, x_{i}$
are no smaller than other $d_{\mathsf{row}}(i,j)$, $j=1,\ldots,r_2-1$.
In order to put pixels in different rows together, we need to \emph{combine}
each category of the $i+1$-th row \emph{with} a suitable category of the $i$-th row \footnote{The phrase ``\emph{combine} a category $A$ \emph{with} a category $B$" is referred to as setting $B\leftarrow A\cup B$ and $A\leftarrow\varnothing$.},
$1\leq i\leq r_1-1$. To do this, for each $1\leq j\leq r_2$, we define another category sequence $J_{k,j}$, $k=1,\ldots,y_j+1$ as follows:
\[ \!\!  \underbrace{(1,j),\!\cdots\!,(i_1,j)}_{J_{1,j}} \underbrace{(i_1\!+\!1,j),\!\cdots\!,(i_2,j)}_{J_{2,j}} \cdots
 \underbrace{(i_{y_j}\!+\!1,j),\!\cdots\!,(r_1,j)}_{J_{y_j+1,j}}  \]
where $i_1<i_2<\cdots<i_{y_j}$ are chosen to be the indices such that $d_{\mathsf{col}}(i_k,j)$, $k=1,2,\ldots, x_{i}$
are no smaller than other $d_{\mathsf{col}}(i,j)$, $i=1,\ldots,r_1-1$.
Next we shall determine which category of $I_{i,1}, I_{i,2},\ldots, I_{i, x_{i}}$
be the category $I_{i+1,t}$ combined with. We define a category distance metric between $I_{i,k}$ and $I_{i+1,t}$ by
\begin{equation}\label{eq-Dalpha}
 \!\! D( I_{i,k},  I_{i+\!1,t})\!=\!\left\{
 \begin{array}{ll} +\infty & \mbox{~if~Condition~1~or~2~holds} \\
\sum\limits_{j\in\Xi_{k,t}}\!\frac{d_{\mathsf{col}}(i,j)}{|\Xi_{k,t}|} & \mbox{~otherwise}
   \end{array}
 \right.
\end{equation}
where
\begin{description}
\item[Condition 1:]  $\nexists~ j$  such that $(i,j)\in  I_{i,k}, (i+1, j)\in  I_{i+1,t}$

\item[Condition 2:] $\exists~ j$    such that $(i,j)\in  I_{i,k}$, $(i+\!1,j)\in  I_{i+\!1,t}$,
$~~~~~~~~~~~~~~~~\{(i,j),(i+1,j)\}\not\subseteq  J_{l,j}, ~\forall~ l= 1,\ldots, y_{j}$
\end{description}
and where
\begin{align*}
&\Xi_{k,t}:=  \big\{1\leq j\leq r_2:\exists~ 1\leq l\leq y_{j} \mbox{~such that~} (i,j)\in  I_{i,k}\\
&~~~~~~~~~~~(i+1,j)\in  I_{i+1,t}, \{(i,j), (i+1,j)\} \subseteq  J_{l,j}  \big\}
\end{align*}
Condition 1 indicates that two categories $I_{i,k}$, $I_{i+1,t}$ share no common column index;
Condition 2 indicates that $I_{i,k}$, $I_{i+1,t}$ share common column indices but those indices are located on the boundary between two groups
of similar-value pixels in columns. 
If $I_{i,k}$,
$I_{i+1,t}$ satisfy either of two conditions for all $k$, then $I_{i+1,t}$ is not combined with any categories; otherwise, we set $k'\leftarrow\arg\min_k D( I_{i,k},  I_{i+1,t})$ and combine $I_{i+1,t}$ with $I_{i,k'}$, where $D( I_{i,k},  I_{i+1,t})$ is given by the average of all color distance metrics in $j$-th column direction such that $(i,j)$, $(i+1,j)$ belong to a common category $J_{t,j}$. We illustrate the \textsf{cluster} phase in Fig.~\ref{fig-cluster} and we give details in Section~C of the supplementary material.

\section{Mesh Compression using CGIMs}\label{sec-6}

\begin{table*}[t]
\centering
\caption{Comparisons of resolutions, file sizes and running time of  CGIMs with respect to traditional GIMs for eight models. The first to fourth columns are the names, number of vertices, number of faces, file sizes of original models, respectively; the seventh to tenth columns are running time of CGIMs, modified CGIMs, lossless reconstruction from CGIMs, lossy reconstruction from CGIMs (averaged with five compression rates in JPEG2000), respectively; the thirteenth to fifteenth columns are running time of traditional GIMs, lossless reconstruction from GIMs, lossy reconstruction from GIMs (averaged with five compression rates in JPEG2000), respectively.}\label{tab-igim}
\renewcommand{\multirowsetup}{\centering}
\begin{tabular}{m{8.5mm}|m{6mm}|m{6.5mm}|m{4mm}!{\vrule width1.0pt} m{14mm}|m{3mm}|m{5.5mm}|m{8.5mm}|m{7mm}|m{7mm}
!{\vrule width1.0pt} m{14mm}|m{3mm}|m{3mm}|m{7mm}|m{7mm}}
\hline
\multirow{3}{8.5mm}{\!models} &\multirow{3}{6mm}{$|V|$} & \multirow{3}{6mm}{$~|F|$} & \multirow{3}{4mm}{size \\ $\!\!\!\mbox{(MB)}$} &   \multicolumn{6}{c|}{CGIM resolutions/size (MB)/time (s)}    &\multicolumn{5}{c}{GIM resolutions/size (MB)/time (s)} \\
\cline{5-15}
&&&  &\multirow{2}{12.6mm}{$\!\!\mbox{resolutions}$} &\multirow{2}{3mm}{$\!\mbox{size}$}& \multirow{2}{5.5mm}{\!\!\!CGIM} & $\!\!\!\mbox{modified}$  & $\!\!\mbox{lossless}$ & $\!\!\mbox{lossy}$ & \multirow{2}{12.6mm}{$\!\mbox{resolutions}$} & \multirow{2}{3mm}{$\!\mbox{size}$} &\multirow{2}{3mm}{\!\!\!GIM} & $\!\!\mbox{lossless}$  & $\!\!\mbox{lossy}$
\\
&&&  &                                               &             &                     & $\!\!\mbox{CGIM}$       & $\!\!\mbox{reconst}$ & $\!\!\mbox{reconst}$ &                                            &                              & & $\!\!\mbox{reconst}$   & $\!\!\mbox{reconst}$ \\
\Xhline{0.6pt}
$\!\!\!cylinder$  & 1663 &  3299  & 0.10 &$218\times 294$  & $\!\!0.18$ & 20 &  69   &     4   &  87         & $512\times 512$ & $\!\!0.77$ & 56       &   20   & 20 \\
\hline
$fist$            &1198 &  2369 & 0.07 & $122\times 197$   &$\!\!0.07$  & 10 &  35   &   2     &  23         & $512\times 512$ & $\!\!0.77$ & 34       &   20   &  20 \\
\hline
$foot$    & $\!10010$  & $\!19974$  & 0.66 & $423\times 970$ &$\!\!1.17$& 667  & 1351  & 31    & 690     & $512\times 512$ & $\!\!0.77$ & $\!126$      &  21    &  21 \\
\hline
 $camel$  &$\!11381$ & $\!22704$  & 0.75 & $475\times 897$  &$\!\!1.21$& 758 & 1716 &  32   &   905   & $512\times 512$ & $\!\!0.77$ & $\!184$      & 21     &  21 \\
\hline
$face$     &$\!40969$  & $\!81485$ & 2.74  & $538\times\! 1493$ &$\!\!2.29$& 902  & 2294  &  67  & {1638}       &$\!\!1024\times\! 1024$ & $\!3.0$&  $\!617$  & 83     & 83  \\
\hline
$head$     &$\!25445$ & $\!50801$ & 1.69 &$757\times\! 1664$  &$\!\!3.60$& $\!\!5445$ & $9801$   &  96 & 3259    &$\!\!1024\times\! 1024$& $\!3.0$& $\!311$    &      84     & 83  \\
\hline
$david$    &$\!23889$ & $\!47280$ & 1.53  &$841\times\! 1751$  &$\!\!4.21$& $\!\!7889$ & $\!17280$   &  203 &  5491  &$\!\!1024\times\! 1024$&$\!3.0$ & $\!270$     &   83   &  83\\
\hline
$hand$     &$\!53054$ & $\!\!105860$ & 3.71 &$\!\!1468\times\! 3048$ &$\!\!12.8$& $\!\!13054$ & $\!20866$   & 338  & $\!12872$  &$\!\!1024\times\! 1024$& $\!3.0$& $\!922$ & 84 & 83\\
\hline
\end{tabular}
\end{table*}
This section gives experimental results of CGIM algorithms and results of CGIM-based mesh compression.
We run all experiments  on an i7-2600 3.4GHz machine with 16GB RAM using Matlab R2013a, and we use Algorithm~\ref{alg-modifyiso} for the \textsf{isomatrix} phase,
where the threshold parameters are taken as $\alpha=5$.
We show the CGIMs, CNIMs (connectivity-preserving normal images, each pixel of which corresponds to the normal of an element in the V-matrix) and traditional GIMs, NIMs for a number of mesh models in Fig.~\ref{fig-igim2}, including four open genus-zero meshes ($cylinder$, $fist$, $face$, $david$) and two closed genus-zero meshes ($bunny$, $venus$).
We show the comparisons of resolutions, file sizes and running time of CGIMs with respect to traditional GIMs in Table~\ref{tab-igim}.
As we see, the file sizes of CGIMs  of $foot, head, david, hand$ are greater, and the sizes of CGIMs of $fist,face$ are relatively small compared with the sizes of original models.
The reason is that those models  contain toes, fingers or other details and hence have large geometric distortions, while $fist, face$ both have lower distortions.
Moreover,  Table~\ref{tab-igim} shows that CGIMs have a much longer running time compared with traditional GIMs, especially for models with large sizes. This is because CGIMs spend much time on obtaining the V-matrix connectivity-preserving to the mesh (Algorithm~\ref{alg-modifyiso}) and reconstructing meshes (the \textsf{cluster} phase). We admit the less efficiency of CGIMs as such two phases of CGIMs are unavoidable.

We implement the lossy compression by using CGIMs and traditional GIMs for six models: $cylinder$, $fist$, $david$, $face$, $bunny$, $venus$ \footnote{The resolutions of CGIMs for representing $bunny$, $venus$ are $602\times 1991$, $1068\times 2706$ respectively.}. For traditional GIMs, we choose two resolutions ($256\times 256$, $512\times 512$ for $cylinder$ and $fist$ as they have relatively smaller sizes; $512\times 512$, $1024\times 1024$ for the other four models) of 8-bit images with JPEG2000 as image codec, where the wavelet kernel is W9X7 and the wavelet transform level is five; for CGIMs, we choose
JPEG2000 as image codec (with the same wavelet kernel and transform level) and the MBCIM  as image codec, respectively. The MBCIM scheme  was proposed  for  compressing compound images, many pixels of which share common values, which are close to CGIM arrays. The MBCIM scheme is introduced briefly in Section~D of the supplementary material.

\begin{figure*}[htbp] \centering
\includegraphics[width=1\textwidth]{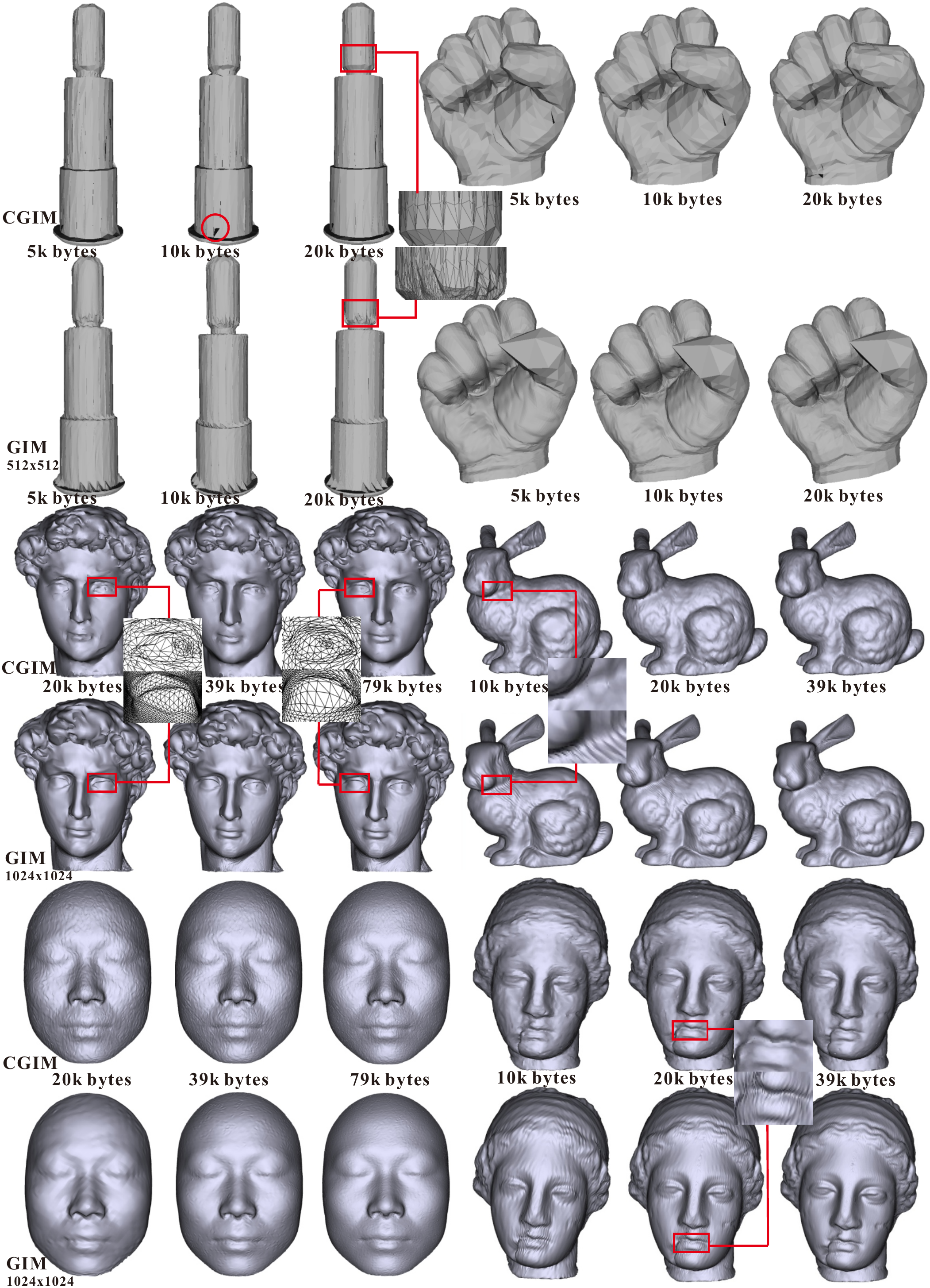}
\parbox{16cm}{\caption{Reconstructed surfaces from compressed CGIMs (odd rows) vs compressed GIMs (even rows) with JPEG2000. Six models: $cylinder, fist$ (rows one, two),
$david, bunny$ (rows three, four), $face, venus$ (rows five, six). } \label{fig-igim3}}
\end{figure*}

\textbf{Qualitative results}\quad
We show mesh reconstructions from lossy compressed CGIMs in Fig.~\ref{fig-igim3}. CGIMs recover the thumb part of the $fist$ model better than GIMs do; this is mainly because the parametrization puts the original vertices of the thumb into a smaller region and imposes an inadequate sampling over that region.
Reconstructed meshes for GIMs look smooth in general, but fail in representing some details (e.g., $cylinder$'s top and $david$'s eyes); nevertheless, reconstructed meshes for CGIMs look coarse in lower target rates (e.g., $bunny$'s ears).

\begin{figure*}[htbp] \centering
\includegraphics[width=0.75\textwidth]{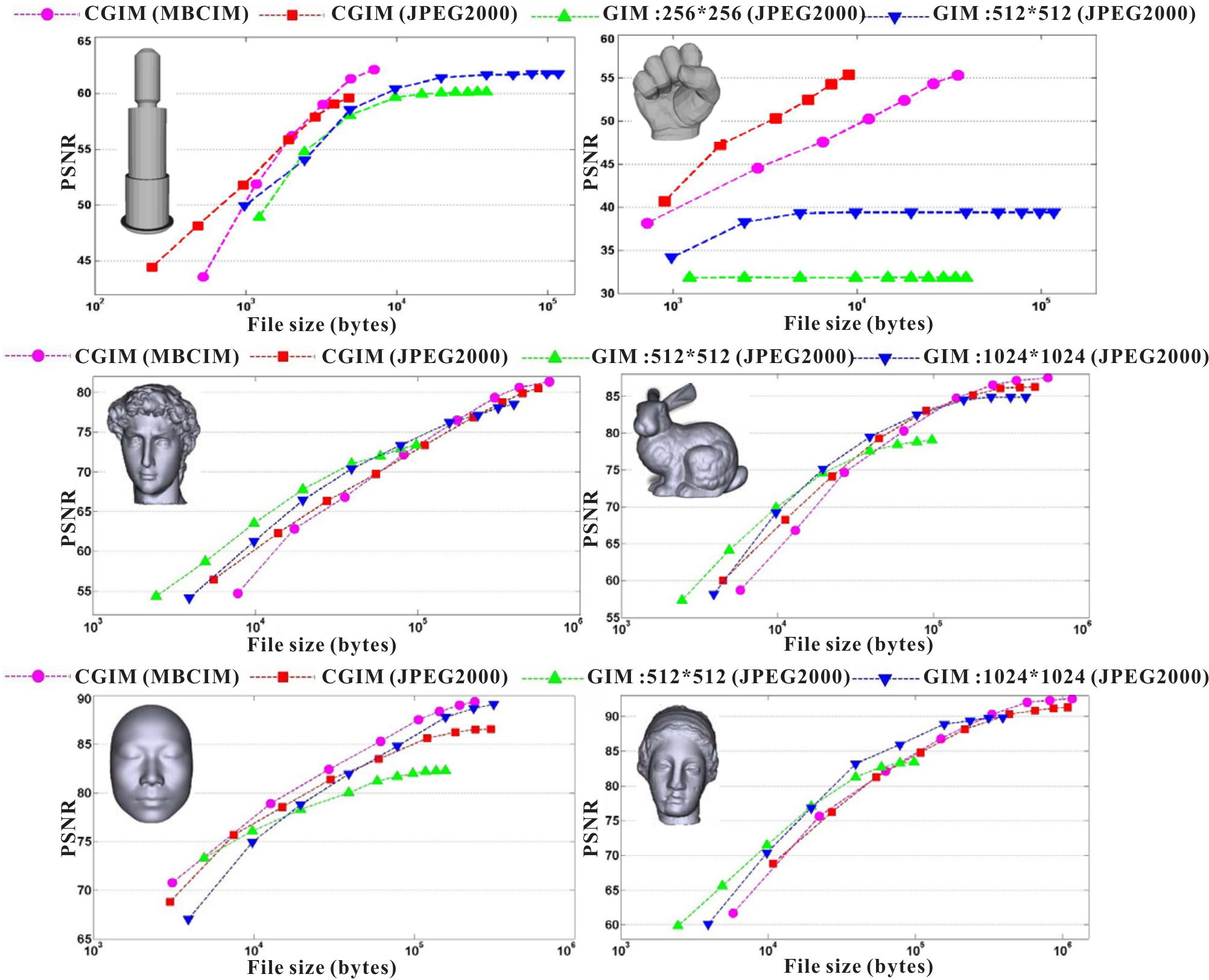}
\caption{Rate distortions for mesh reconstruction from CGIMs with MBCIM (cycles), CGIMs with JPEG2000 (cubes),
GIMs of smaller resolutions  with JPEG2000 (triangles; $256\times 256$ for $cylinder,fist$ and $512\times 512$ for $david,bunny,face,venus$), and GIMs of larger resolutions with JPEG2000 (inverted triangles; $512\times 512$ for $cylinder,fist$ and $1024\times 1024$ for $david,bunny,face,venus$).} \label{fig-psnr}
\end{figure*}

\textbf{Quantitative results}\quad
We show the PSNR curves in Fig.~\ref{fig-psnr}, where the $x$-axis represents file sizes given by setting different target rates (in JPEG2000)  or different quantization parameters (in MBCIM), and the $y$-axis represents the PSNR values.
For $cylinder$, $face$, $fist$, CGIMs achieve better than GIMs in the whole part of curve, which is mainly because those three models have small resolutions with respect to the original size of the mesh. In  particular, $fist$ has a bad parametrization on its thumb part, leading to an insufficient sampling over that part and hence a bad reconstruction using GIMs. On the other hand, GIMs with resolutions $1024\times 1024$ achieve better results for $david$, $bunny$, $venus$ in the most part of curves, while CGIMs
achieve better results when the compression rate tends to be lower. Moreover we see that CGIMs with MBCIM coding achieve better rate distortions than CGIMs with JPEG2000. Also, the rate distortions strongly depend on the resolutions of CGIMs: the smaller the resolutions are, the better rate distortions CGIMs achieve.

Therefore, reconstruction errors of CGIMs for lossy compression depends on the quantization errors of coordinates of vertices,  incorrect cluster algorithms during reconstructions and the incorrect connectivity incurred by quantization errors and incorrect cluster algorithms; moreover, the PSNR depends on the three ingredients given above, as well as the CGIM resolutions.

\textbf{Discussion}\quad  Fig.~\ref{fig-igim3} implies that CGIMs have promising advantages in preserving details compared with traditional GIMs. The reason is that GIMs
treat a mesh with a regular sampling over parametrized domain and hence incur a loss in high distortion parts (e.g., $david$'s eyes), whereas CGIMs treat details and non-details of a mesh evenly as the elements of the V-matrix. Although the magnified details show that GIMs produce more compact grids in $david$'s eyebrows than CGIMs do, what CGIMs show is close to the faithful geometry of the model.

\begin{figure}[htbp] \centering
\includegraphics[width=0.34\textwidth]{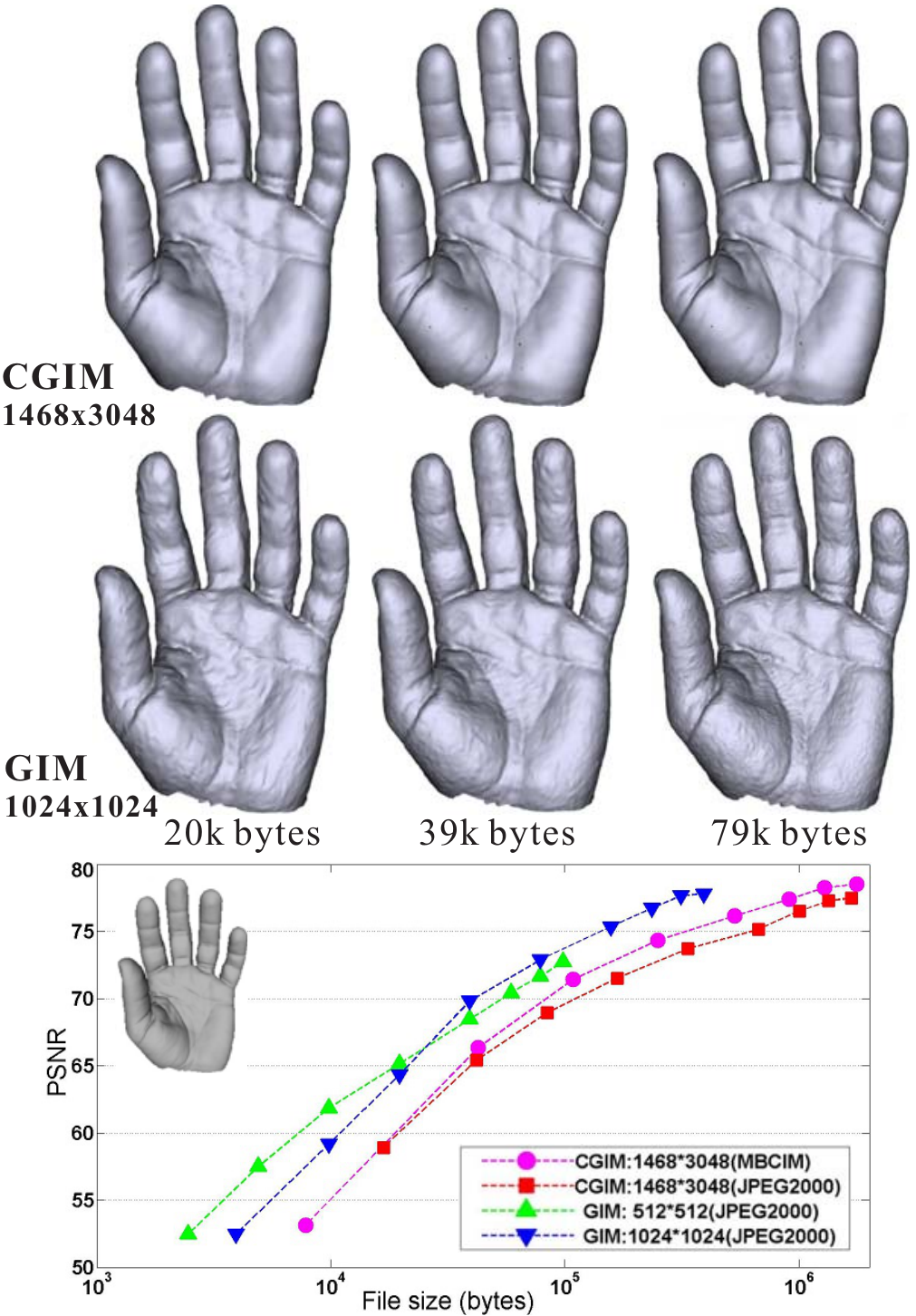}
\caption{A failure example whose rate-distortion curve for CGIMs is lower than the one for GIMs.
This is because of its large distortion (with five fingers) leading to blow-up resolutions of CGIMs,
which needs higher compression rate than traditional GIMs under the same compressed data size.} \label{fig-failure}
\end{figure}

Moreover, with the same codec JPEG2000, CGIMs do not perform better PSNRs than GIMs, but CGIMs with MBCIM codec perform better PSNRs than CGIMs with JPEG2000. This is mainly because CGIMs usually have large resolutions compared with GIMs as we mentioned before, which increases the compressed file size for the same target rate during encoding. For large distortion models ($david$, $bunny$, $venus$), CGIMs do not have better PSNRs for small target rates, as the quantization error produces incorrect connectivity in the \textsf{cluster} algorithm. Such an error decreases and accordingly the PSNR increases when the target rate is larger. Also, MBCIM is more suitable than JPEG2000 for CGIM compression according to the experimental results. Yet, CGIMs give bad PSNRs for meshes with much larger distortion (see Fig.~\ref{fig-failure}, a failure example of $hand$), as those meshes have large resolutions of CGIMs and consequently the compressed file size is large for a given compression rate.

\section{Conclusions}\label{sec-7}
\textbf{Advantages}\quad  This paper proposes  CGIMs, which represent genus-zero triangular meshes by embedding a mesh onto a rectangular array which preserves the connectivity of vertices with round-off errors on coordinates of vertices. Compared with traditional GIMs, the advantages of CGIMs mainly include:

(i) A minimum reconstruction error estimate with respect to the original mesh.

(ii) A highly precise reconstruction of detailed or large distortion parts of meshes during CGIM-based mesh compression.

(iii) A parametrization-free CGIM algorithm which avoids both solving a high-dimensional optimization problem
and computing the coordinates of sampling points.

\textbf{Limitations}\quad  We conclude the limitations of CGIMs as follows.

(i) The greatest disadvantage of CGIMs is that CGIMs have much redundancy especially for meshes with large distortion, which makes CGIMs need higher compression rate than traditional GIMs under the same limit of data size. This can be seen from Fig.~\ref{fig-igim3} where CGIMs show less smoothness on the models  $face, venus$.

(ii) CGIMs are less efficient than traditional GIMs. According to Table~\ref{tab-igim}, CGIMs spend much more time on obtaining the V-matrix and reconstructing meshes (the \textsf{cluster} phase) than traditional GIMs.

(iii) Unlike GIMs, traditional image codecs are not suitable for CGIMs because most pixels of CGIMs share the same value and such a structure brings much trouble during the reconstruction of lossy compression. To overcome this shortcoming, other image codec for CGIMs must be developed.

(iv) Under high compression rate, the \textsf{cluster} phase may result in incorrect connectivity of vertices or orientation of faces (see the dark face at the bottom of $cylinder$ in row one, column two of Fig.~\ref{fig-igim3}), which destroys the connectivity-preserving property and limits the applications of CGIMs to lossy compression.

\textbf{Future work}\quad
In the future, we shall consider the following research problems regarding CGIMs.

(i) Developing multi-chart CGIMs with smaller resolutions. We shall consider first partitioning it into a collection of submeshes, each of which is represented by a CGIM with small resolutions, and then packing all sub-CGIMs into a single CGIM.

(ii) Developing appropriate codecs for CGIMs. Because many pixels of CGIMs share a common value, it is rewarding to consider more efficient codec for CGIMs such as run-length coding.

(iii) Applying CGIMs to hierarchical mesh compression. We shall consider GIMs which preserve details, features or region of interest of meshes by using CGIM patches, and impose ROI codings on such GIMs.

%

\end{document}